\documentclass[10pt,letterpaper]{article}
\usepackage[top=0.85in,left=2.75in,footskip=0.75in]{geometry}

\usepackage{changepage}

\usepackage[utf8]{inputenc}

\usepackage{textcomp,marvosym}

\usepackage{fixltx2e}

\usepackage{amsmath,amssymb}

\usepackage{cite}
\usepackage{color}
\usepackage{nameref,hyperref}

\usepackage[right]{lineno}

\usepackage{microtype}
\DisableLigatures[f]{encoding = *, family = * }

\usepackage{rotating}

\usepackage{setspace} 

\usepackage[aboveskip=1pt,labelfont=bf,labelsep=period,justification=raggedright,singlelinecheck=off]{caption}

\makeatletter
\renewcommand{\@biblabel}[1]{\quad#1.}
\makeatother

\date{}

\usepackage{lastpage,fancyhdr,graphicx}
\usepackage{epstopdf}
\fancyheadoffset[L]{2.25in}
\fancyfootoffset[L]{2.25in}

\def\ff{\mathcal{F}}   
   
\def\rpi{R_\Pi}
\def\MPa{\; \text{\it MPa}}

\begin{document}
\vspace*{0.35in}

\begin{flushleft}
{\Large
\textbf\newline{Membrane Mechanics of Endocytosis in Cells with Turgor}
}
\newline
\\
Serge Dmitrieff and François Nédélec*\\
\bigskip
Cell Biology and Biophysics Unit, European Molecular Biology laboratory, Meyerhofstrasse 1, 69123 Heidelberg, Germany.
\\
\bigskip

* Corresponding author : nedelec@embl.de

\end{flushleft}

\section*{Abstract}


Endocytosis is an essential process by which cells internalize a piece of plasma membrane and material from the outside. 
In cells with turgor, pressure opposes membrane deformations, and increases the amount of force that has to be generated by the endocytic machinery. To determine this force, and calculate the shape of the membrane, we used physical theory to model an elastic surface under pressure. Accurate fits of experimental profiles are obtained assuming that the coated membrane is highly rigid and preferentially curved at the endocytic site. The forces required from the actin machinery peaks at the onset of deformation, indicating that once invagination has been initiated, endocytosis is unlikely to stall before completion. Coat proteins do not lower the initiation force but may affect the process by the curvature they induce. In the presence of isotropic curvature inducers, pulling the tip of the invagination can trigger the formation of a neck at the base of the invagination. Hence direct neck constriction by actin may not be required, while its pulling role is essential.  Finally, the theory shows that anisotropic curvature effectors stabilize membrane invaginations, and the loss of  crescent-shaped BAR domain proteins such as Rvs167 could therefore trigger membrane scission.

\section*{Author Summary}

Cells use endocytosis to intake molecules and to recycle components of their membrane. Even in its simplest form, endocytosis involves a large number of proteins with often redundant functions that are organized into a microscopic force-producing ``machine". Understanding how much force is needed to induce a membrane invagination is essential to understand how this endocytic machine may operate. We show that experimental membrane shapes are well described theoretically by a thin sheet elastic model including a difference of pressure across the membrane due to turgor. This allows us to integrate the different contributions that shape the membrane, and to compute the forces opposing membrane deformation. This calculation provides an estimate of the pulling force that must be generated by the endocytic machine. We also identify a membrane instability that could lead to vesicle budding.


\section*{Introduction}

Endocytosis enables cells to internalize extracellular material and to recycle membrane components \cite{mukherjee1997endocytosis}. During this process, the plasma membrane is deformed into an invagination progressing inwards, which is severed and eventually released in the cytoplasm as a vesicle. Key endocytic components have been identified in several systems. This process usually involve membrane coating proteins (such as clathrin), their adaptors (such as epsins) and actin microfilaments together with associated factors \cite{mcmahon2011molecular}. We focused on the yeast model system, in which endocytosis is well characterized experimentally.
The abundance and localization of the principal proteinaceous components have been measured as a function of time both in {\it S. pombe} \cite{sirotkin2010quantitative, berro2014} and {\it S. cerevisiae} \cite{kaksonen2005modular,picco2015visualizing}. To understand how these components work together to deform the membrane, it is necessary to consider the physical constraints under which the task is performed {\em in vivo}. 
While in animal cells, invaginations are opposed mostly by membrane tension and elasticity \cite{ewers:2009}, turgor pressure strongly  opposes invaginations in plants and fungi. In those cells, the difference of osmolarity with the outside cause a large pressure pushing membrane against the cell wall.


\begin{figure}[t]   
\begin{adjustwidth}{-2.25cm}{0cm}
\centering
\includegraphics[width=7.5cm]{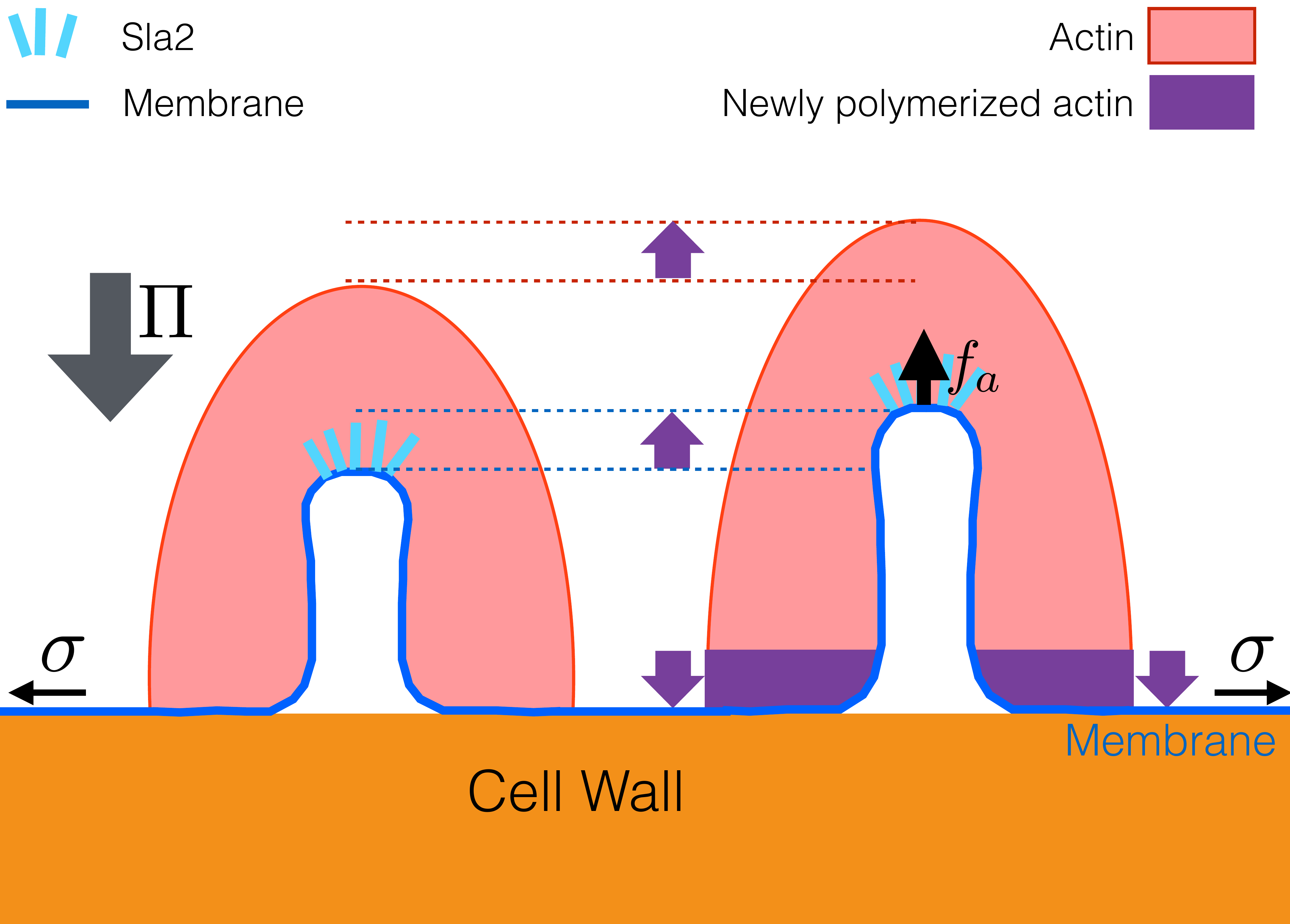} \hspace{0.5cm} \includegraphics[width=5cm]{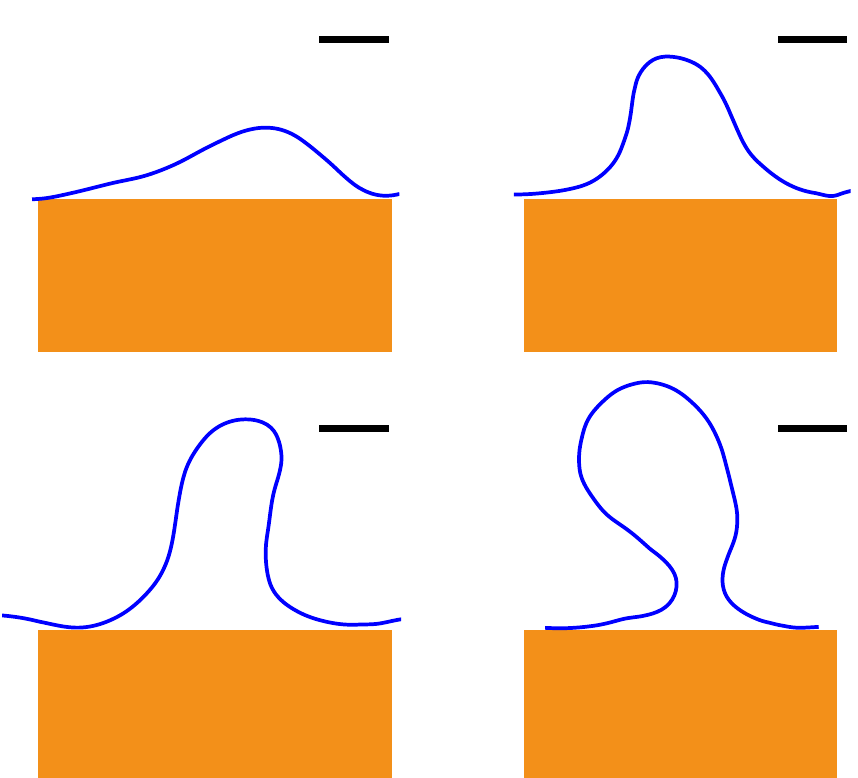}
\caption{\label{schematic_prof}
Left : schematic representation of endocytosis. The membrane is connected mechanically to the actin network through sla2. Actin is polymerizing close to the basal membrane, thus generating an upwards (pulling) force $f_a$. This force overcomes the rigidity $\kappa$ and tension $\sigma$ and the turgor pressure $\Pi$ that all oppose invaginations. Right : four experimental membrane profiles at different stages, measured in an electron-microscopy study \cite{kukulski2012plasma}. Scale bar : $20nm$.}
\end{adjustwidth}
\end{figure}

The magnitude of the pressure has been measured in different walled cells using various methods. For plant cells, studies converge to the range $0.2 \text{---} 1 \MPa$ (see \cite{beauzamy2014flowers} for a careful review of the subject). In the yeast {\it S. pombe}, an effective pressure of $0.85 \pm 0.15  \MPa$ was derived from studying the buckling of the rod-like cells in micro fabricated chambers \cite{minc2009mechanical}. Based on the variation of volumes upon changes in osmolarity, the pressure in {\it S. cerevisiae} was recently estimated to be $0.6 \pm 0.2 \MPa$ \cite{schaber2010biophysical}, while an older study concluded $0.2 \MPa$ for stationary phase cells \cite{de1996passive}.
It has been suggested that the pressure could be decreased locally by releasing osmolytes at the endocytic patch \cite{carlsson2014force}. This would however only lower the {\em osmotic} pressure, which is not opposing endocytosis. The {\em hydrostatic} pressure gradients equilibrate at the speed of sound (possibly $1500\; m/s$), and given the size of yeast, this is considerably faster than an endocytic event ($\sim 5 s$). 

The pressure, thus of the order of $1\MPa \sim 1\; pN/\mbox{nm}^2$, pushes the plasma membrane outward uniformly. This effect is balanced by an equal force from the cell wall, wherever the membrane is in contact with the cell wall (Fig. \ref{schematic_prof}, left). The pressure strongly opposes endocytic membrane invagination, as membrane and cell wall must come apart. The large required force is produced in yeast by an actin machinery, which forms a crosslinked network of rigid filaments around the invagination (see \cite{kukulski2012plasma} and Fig. \ref{schematic_prof}, left).
Actin polymerizes close to the basal plasma membrane \cite{picco2015visualizing}. The newly inserted F-actin at the bottom of the network is though to lift the entire network away from the cell wall (upward on Fig. \ref{schematic_prof}, left)\cite{carlsson2014force}. 
For this force to be productive, the actin network should be attached to the tip of the invagination, and this is the role of the protein sla2, which is required for endocytosis - while several other proteins, including clathrin, are dispensable \cite{kaksonen2006harnessing,kaksonen2005modular}. 
This essential role for actin is further supported by the fact that actin assembly precedes or coincides with membrane deformation \cite{kukulski2012plasma,picco2015visualizing}. The interplay between actin and pressure was nicely demonstrated by showing that impairing the arp2/3 complex (an actin nucleator) delayed the invagination, while decreasing the hydrostatic pressure (by adding sorbitol to the media) had the opposite effect \cite{basu2014role}.

Coat and associated proteins play a  role in endocytosis, notably by recruiting the endocytic machinery, and keeping the actin network physically connected to the membrane. Proteins that bind to the lipid bilayer can also directly induce the membrane to curve \cite{mcmahon:2005,zimmerberg2006proteins}. Two familiar examples in yeast endocytosis are clathrin \cite{mukherjee1997endocytosis} and Rvs167 \cite{wigge1998amphiphysin}. The induced curvature is expected to be qualitatively different for these two proteins, because clathrin proteins form regular triskelion \cite{kirchhausen1986configuration} while Rvs167 is shaped as a crescent \cite{peter2004bar}. Moreover, clathrin adaptors can bind to membrane, cargo, actin and/or clathrin and have essential functions in endocytosis \cite{reider2011endocytic}. Additionally, some adaptor proteins can induce membrane curvature \cite{ford2002curvature}.

The shape of the membrane can be predicted by minimizing an effective deformation energy \cite{helfrich:1973}.  This approach has been used successfully in different systems \cite{deuling1976curvature,wiese1992budding,derenyi2002formation}, but the case of yeast endocytosis where the membrane detaches from the cell wall despite a large hydrostatic pressure has not been analyzed theoretically to our knowledge. Most previous work focused on the endocytosis of viruses in animal cells, which occurs by wrapping of the membrane around viral particles in the absence of hydrostatic pressure \cite{gao2005mechanics,nowak2008membrane,ewers:2009} ; the constriction of vesicle necks was also studied recently \cite{bovzivc2014direct}, but without comparison to physiological results. Yeast endocytosis was modelled before, mostly without turgor \cite{agrawal2010minimal,Zhang2015508}.
One study considered a difference of pressure across the plasma membrane \cite{walani2015endocytic}, but adopted a value of the pressure estimated in yeast spheroplasts \cite{gustin1988mechanosensitive}. The pressure in these cells, which lack a cell wall, is at least two orders of magnitude lower than under more physiological conditions \cite{minc2009mechanical}, and the forces in this study are consequently severely underestimated.

In this work, we integrate the contributions of pressure, coat proteins and membrane properties during the endocytic invagination, providing an estimate of the force that actin must exert to induce the invagination. We study the effect of coat proteins such as clathrin that induce isotropic curvature, and contrast it with crescent-shape proteins such as BAR-domain that induce curvature only in one direction. Finally, we discuss whether the combination of actin-mediated pulling together with the removal of Rvs167 is sufficient to lead to vesicle internalization.

\section*{Methods}

We predict membrane shapes by minimizing a deformation energy, using a Helfrich-type Hamiltonian \cite{deuling1976curvature,julicher1994shape}, in which membrane deformations are penalized by a bending rigidity $\kappa$ and tension $\sigma$. 
We write $\Pi$ the difference of hydrostatic pressure between the inside and the outside of the cell. 
In addition, we assume a point force $f_a$ pulling the apex of the invagination, to represent the driving force generated by the actin cytoskeleton \cite{picco2015visualizing,carlsson2014force}. 
We thus implicitly assume that the forces produced by actin polymerization at the base of the invagination, and possibly by myosin motors or other processes, are transmitted to the tip of the invagination over the actin network (see Fig. 1). 
We note $\kappa$ the rigidity of the membrane together with its coat of proteins \cite{mcmahon:2005,zimmerberg2006proteins}.
Moreover, we  consider that the coat proteins curve the membrane either by scaffolding or inserting themselves in the membrane. We first describe proteins such as clathrin, that induce an isotropic curvature $C_0$, with the same radius of curvature in both directions.

To write the membrane deformation energy, we introduce $C = ({\frac{1}{R_1}+\frac{1}{R_2}})$ the local curvature, $V$ the volume inside the invagination, $S$ the surface area of the membrane that is not in contact with the wall, and $L$ the height of the invagination.  The total energy $\ff$ of an invagination reads \cite{derenyi2002formation}:
\begin{eqnarray}
\ff = \iint_S \left[ \frac{\kappa}{2} \left(C-C_0\right)^2  + \sigma \right] dS +  \Pi V  - f_a L\, .
\label{generalff2}
\end{eqnarray}

The rigidity term will tend to make invaginations as large as possible to minimize their curvature, while both pressure and tension will tend to make invaginations smaller, to minimize their volume and surface, respectively. 
Therefore an invagination dominated by pressure and rigidity will have a typical width $\rpi = \sqrt[3]{\kappa/2\Pi}$ while an invagination dominated by tension and rigidity will have a typical width $\lambda \sim \sqrt{\kappa / 2 \sigma}$.

From measured values of $\Pi, \;\sigma$ (Table \ref{characteristics}), and using the rigidity of a naked membrane ($\kappa \sim 40\; k_bT$), we find $\rpi \sim 3\; nm $ and  $\lambda > 9\; nm$. Since $\rpi < \lambda$ in this case, we expect that invaginations will have a typical radius $\rpi$ and tension will not significantly affect the shape of the invagination. In this estimate, we have used a lower bound for $\kappa$ given by the rigidity of a pure lipid bilayer, but the actual value of $\kappa$ will be higher because it should also include the stiffness provided by coat proteins. However, the statement  $\rpi < \lambda$  will be all the more true for higher values of $\kappa$ because $\rpi$ will increase slower than $\lambda$ as $\kappa$ becomes larger. 
We have used an upper bound for the membrane tension $\sigma$, but in reality it could be significantly smaller, since yeast cell have membrane furrows \cite{stradalova2009furrow}, which act as membrane reservoirs and limit the surface tension.  
Moreover, we will see later how fitting the experimental membrane shapes confirms that the contribution of tension is negligible.

\begin{table}[!b]
\caption{
{\bf Physical characteristics of endocytosis in yeast.}}
\begin{tabular}{|l|l|l|l|l|l|l|l|}
\hline
Volume & $V\sim 4 \times 10^4\; nm^3$  & \cite{kukulski2012plasma}  \\
Surface & $S\sim 6 \times 10^3\; nm^2$ & \cite{kukulski2012plasma}  \\ 
Height & $L\; \text{up to}\; 120\; nm$ & \cite{kukulski2012plasma}  \\ 
Tip radius & $R_t \sim  12\; nm$& \cite{kukulski2012plasma}    \\
Tip curvature & $C\sim 2/R_t \sim  0.167\; nm^{-1}$& \cite{kukulski2012plasma}    \\
Pressure   & $\Pi \sim 0.2 \text{---} 1 \; pN.nm^{-2} \qquad$ &\cite{schaber2010biophysical,de1996passive} \\
Tension   & $\sigma < 10^{-3}\; N.m^{-1} \qquad$  &\cite{kwok1981thermoelasticity,morris2001cell} \\
Rigidity   & $\kappa \ge 40\; k_B T\; \text{i.e.}\; 160\; pN.nm $ &\cite{evans:1990}\\
Duration  & $\sim 5\; s$ &\cite{berro2014, picco2015visualizing}\\ \hline
\end{tabular}
\label{characteristics}
\end{table}



Since the competition of pressure with membrane rigidity should control the shape of the invagination, the forces due to pressure and rigidity will have the same order of magnitude and will dominate those due to membrane tension. The total force exerted by pressure over the invagination should be $\Pi S_0$, where $S_0$ is the surface area of the wall that is not in contact with the membrane. 
This area is difficult to delineate in the electron micrographs \cite{kukulski2012plasma}, but using the dimension at the tip of the invagination $R_t$ (Table \ref{characteristics}) to obtain an ersatz $\pi R_t^2$, we estimate the magnitude of the force to $\sim 300\;pN$.

\begin{figure}[t]
\begin{adjustwidth}{-2cm}{0cm}
 \includegraphics[width=17cm]{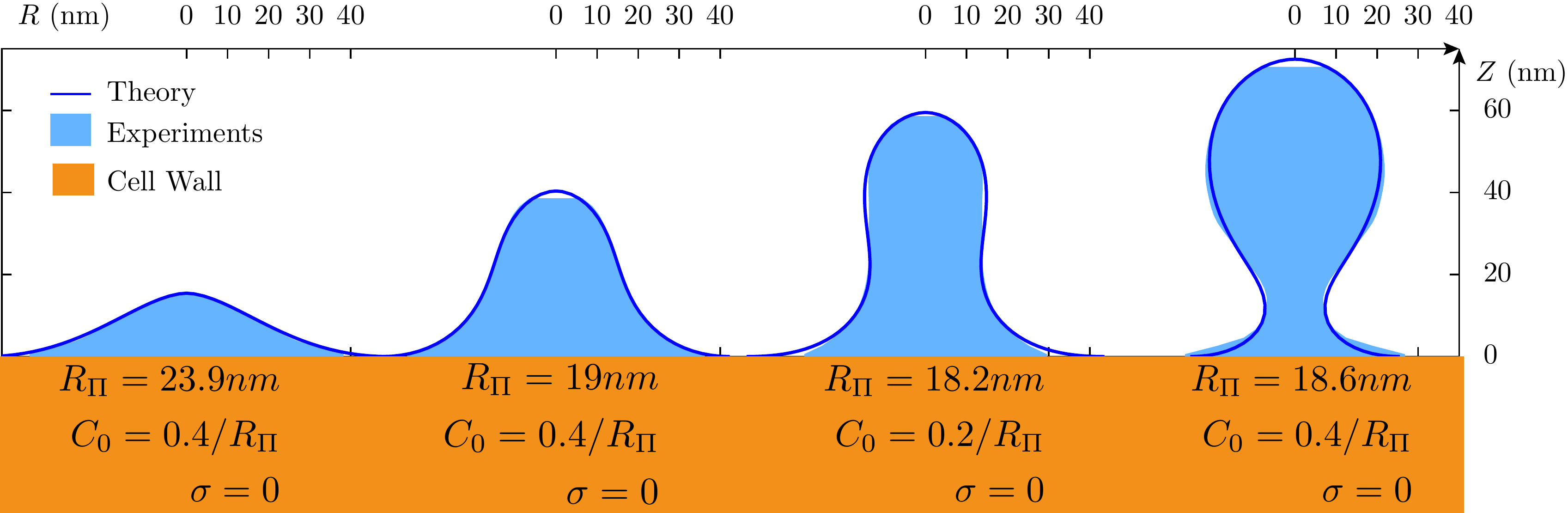}
\caption{\label{profiles}
Predicted membrane profiles (solid lines) and rectified experimental profiles (light blue fill) measured on electron tomograms \cite{kukulski2012plasma}. Three parameters were varied to obtain the best fit: the scale $\rpi \sim 15-25 nm$, the surface tension $\sigma$ and the spontaneous curvature $C_0$. Usually a range of parameter values (including $\sigma > 0$) is suitable to fit each profile. The experimental membrane profiles (Fig. \ref{schematic_prof}, right) have been here straightened to an axisymmetric shape. Details of the method are found in the supplementary information (S.I. 1.5).}
\end{adjustwidth}
\end{figure}

Given this force, and a membrane viscosity $\eta_m \sim 10^{-8}\; kg. s^{-1}$ \cite{cicuta2007diffusion}, we can calculate the time needed to pull a membrane tube over a distance $L\sim 100\; nm$ \cite{kukulski2012plasma}. 
Because the resulting time scale  $L \eta / f \sim 10^{-5}\;s$ is considerably shorter than the duration of an endocytic event (a few seconds \cite{kukulski2012plasma}), the membrane has plenty of time to reach a static equilibrium, at any stage of its evolution. 
In this quasi-static regime, we can derive the membrane shape equations from minimizing the energy in eq. \ref{generalff2}, and following earlier work, we also assume that they are axisymmetric for simplicity (see supplementary information S.I. 1.1).

To compare our predictions to experimental membrane profiles, the measured shapes (Fig. \ref{schematic_prof}, right) were projected to an axisymmetric profile (Fig. \ref{profiles}). 
For a given set of $C_0,\, \sigma,\, R_\Pi$, only one value of $f_a$ allows the theoretical profile to match the invagination shape, and therefore three parameters could be varied to fit experimental membrane profiles (see supplementary information S.I. 1.5 for fitting procedure). 
Generally, we found that measured profiles could be fitted precisely (Fig. \ref{profiles}) except for long invaginations ($L>80nm$, see comment in the discussion).
A given experimental profiles can usually be fitted by a range of parameters rather than a unique set, but the values of the fitting parameters are remarkably consistent.  
The tension is negligible in all cases as expected, and the values of $\rpi$ and $\kappa$ can be interpreted and used to calculate the magnitude of the apical driving force $f_a$.
We will usually express lengths as multiples of $\rpi$ and forces in terms of $f_\Pi = 4 \pi \Pi \rpi^2$, since this makes our results independent of the values $\Pi$ and $\kappa$, which are known only within an interval, without any loss of generality.

 At this stage, we emphasize that the values of $\kappa$ and $C_0$ cannot be known {\em a priori} because these parameters represent properties of the coated membrane which is a complex assembly of interacting lipids and proteins, and not a simple lipid bilayer. In the following, we determine the rigidity $\kappa$ and curvature $C_0$ from the fit, assuming them to be constant over the membrane. In reality the coated membrane is expected to be inhomogeneous, but allowing these parameters to vary spatially would drastically increase the parameter space, such that fitting would be both underdetermined and computationally expensive. Moreover, the overall quality of the fits obtained by assuming $\kappa$ and $C_0$ to be homogeneous, does not warrant an increase in the complexity of the theory. We discuss the implications of heterogeneity in the membrane later.

\section*{Results}

\subsection*{The coated membrane is rigid and preferentially curved}

The fit always provided a well determined value for $\rpi$ in the range $15 \text{---} 25\; nm$, and since $\rpi = \sqrt[3]{\kappa/2\Pi}$, this allowed us to estimate the ratio between the bending rigidity and the pressure. Assuming  $\Pi \sim 1 \MPa$ one gets $\kappa \sim 2000\; k_B T$, while a conservative value of the pressure ($0.2 \MPa$) would yield $\kappa \sim 400\; k_B T$.
The coated membrane is therefore much more rigid than a pure lipid bilayer ($\kappa \sim 40\; k_B T$), leading to wider invaginations. 
The lower estimate corresponds to the expected rigidity of clathrin coated membranes {\it in vitro} ($300\; k_B T$, \cite{jin2006measuring}), while the higher value is also realistic, because other proteins besides clathrin are concentrated at the endocytic site \cite{kaksonen2006harnessing}. 
Adaptor proteins could drastically stiffen the coat by interacting with both clathrin and the membrane \cite{ford2002curvature}.
By increasing the effective rigidity, coat proteins enlarge the invaginations, but as a consequence, the overall resisting force is also increased. 
Importantly, our estimate of the minimal force needed to sustain the invagination, $f_a \sim 3000\; pN$ is not based on the parameters that are poorly determined by the fit (such as $\sigma$).

We were also able to determine the value of the spontaneous curvature $C_0 \sim 0.4 / R_\Pi$.
While this parameter is less well determined,  it corresponds to a radius of curvature $2/C_0 \sim 100\; nm$, which is consistent with the known characteristics of the coat proteins.
In particular, purified clathrin form {\it in vitro} spherical cages with an average radius of $35\; nm$ \cite{cheng2007cryo}. 
The conditions  {\it in vivo} are likely to be different however, since more proteins coat the membrane, and the membrane itself can also contribute to $C_0$. 
Note that we have assumed that curvature applied everywhere to the membrane, rather than in a restricted region as can be expected in reality, but this point will be discussed later.

\subsection*{Large forces are required to initiate endocytosis}
\begin{figure}[t]

\begin{adjustwidth}{-2in}{0in}
\centering
\includegraphics[width=16cm]{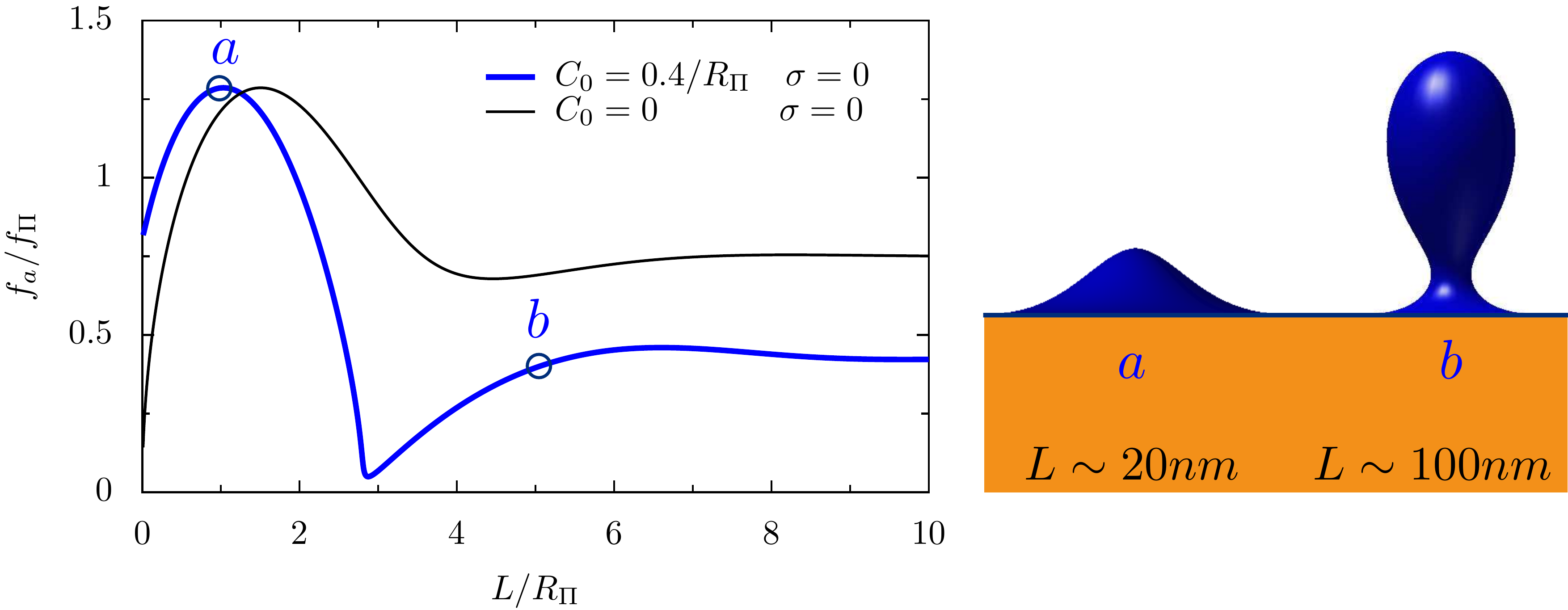} 
\caption{\label{illus_force_length}
Left : pulling force (normalized by $f_\Pi = 2 \pi \kappa / \rpi$) as a function of invagination length (normalized by $R_\Pi$). The blue line corresponds to parameters determined by fitting the experimental profiles, while the black line shows the behavior for an invagination without spontaneous curvature or membrane tension. Derivation of these curves is explained in the supplementary information. Right : membrane profiles for the force maximum ($a$) and for a long invagination ($b$). Lengths are given assuming $R_\Pi \sim 20 nm$.
}
\end{adjustwidth}
\end{figure}

By solving the membrane shape equations for a range of invagination heights $L$, we could compute the required apical force during different stages of the invagination.
While the magnitude of the force is set by $f_\Pi = 4 \pi \Pi \rpi^2$, its variations as a function of $L$ are very instructive.
The force has a non-zero value for $L=0$, for the determined values of the spontaneous curvature $C_0\sim 0.4 / R_\Pi$ (Fig. \ref{illus_force_length}, blue curve).  
$f_a$ increases and reaches a maximum for $L \sim \rpi$, and then decreases to a minimum value for $L \sim 3 \rpi$. For longer invaginations a plateau is reached.
This is very different from membrane without spontaneous curvature (Fig. \ref{illus_force_length}, black line) or for a tension-dominated membrane \cite{derenyi2002formation}, for which the force is minimum for $L = 0$.  
In the presence of spontaneous curvature, the initiation force at $L=0$ is high, the force peaks for smaller invaginations, and the pulling force is generally lowered for longer tubes (compare black and blue lines in Fig. \ref{illus_force_length}). 

The fact that an important initial force $f_0 = f_a(L=0)$ is required to start the invagination is biologically meaningful. 
We estimated $f_0 \sim 0.8 f_\Pi$, corresponding to 60\% of the maximal force. 
Because the membrane is nearly flat if $L \rightarrow 0$, we could obtain analytically $f_0 = 2 f_\Pi \sqrt{\rpi^2 C_0^2 + 2 \sigma R_\Pi / \kappa}$ (see supplementary information S.I. 2.2). 
With $\sigma$ negligible, this expression reduces to $f_0 = 4 \pi \kappa C_0$, which summarizes how coat proteins can affect the initial force, either by changing the spontaneous curvature, or the effective rigidity of the coated membrane.
By itself, spontaneous curvature is not able to lift the membrane off the wall, but rather increases the force needed to start of the invagination. 

Because the force rises sharply with the height of the invagination, actin should be needed to lift the membrane already at the earliest steps of endocytosis, as recently observed \cite{kukulski2012plasma,picco2015visualizing}. 
This idea is consistent with the observation that the delay before significant membrane deformation is observed depends on the competition between actin and pressure \cite{basu2014role}.
It has been suggested that when the membrane is still flat, actin could pull at one site of the membrane while simultaneously pushing on a ring-shaped zone surrounding this site (Fig. \ref{schematic_prof}, left).
This scenario was supported by the recent observation that sla1, an actin organizer, forms rings at endocytic sites on flat membranes, possibly indicating where pushing forces are applied \cite{picco2015visualizing}.
Because the membrane is still juxtaposed to the cell wall at this stage, the pulling and pushing forces generated by the actin network have to be balanced.

Finally, a key feature of the force profile is that the most demanding part of the invagination occurs early during endocytosis ($L < \rpi$).
In mechanical terms, this can lead to a snap-through transition \cite{walani2015endocytic}, in which once a critical threshold is reached, subsequent stages spontaneously follow because they do not require additional efforts.
This may explain that while the duration of the endocytic early phase is highly sensitive to the competition between actin nucleation and hydrostatic pressure, later stages are largely insensitive to pressure \cite{basu2014role}. 
This snap-through force profile also elegantly explains why so few retraction events were observed experimentally (less than $1\%$ of endocytic events fail to complete \cite{kaksonen2005modular}), despite the inherent stochasticity of a biological machine such as endocytosis that only contains a few hundred actin filaments\cite{sirotkin2010quantitative}.

\begin{figure}[t]
\begin{adjustwidth}{-2.25in}{0in}
\centering
\includegraphics[width=18cm]{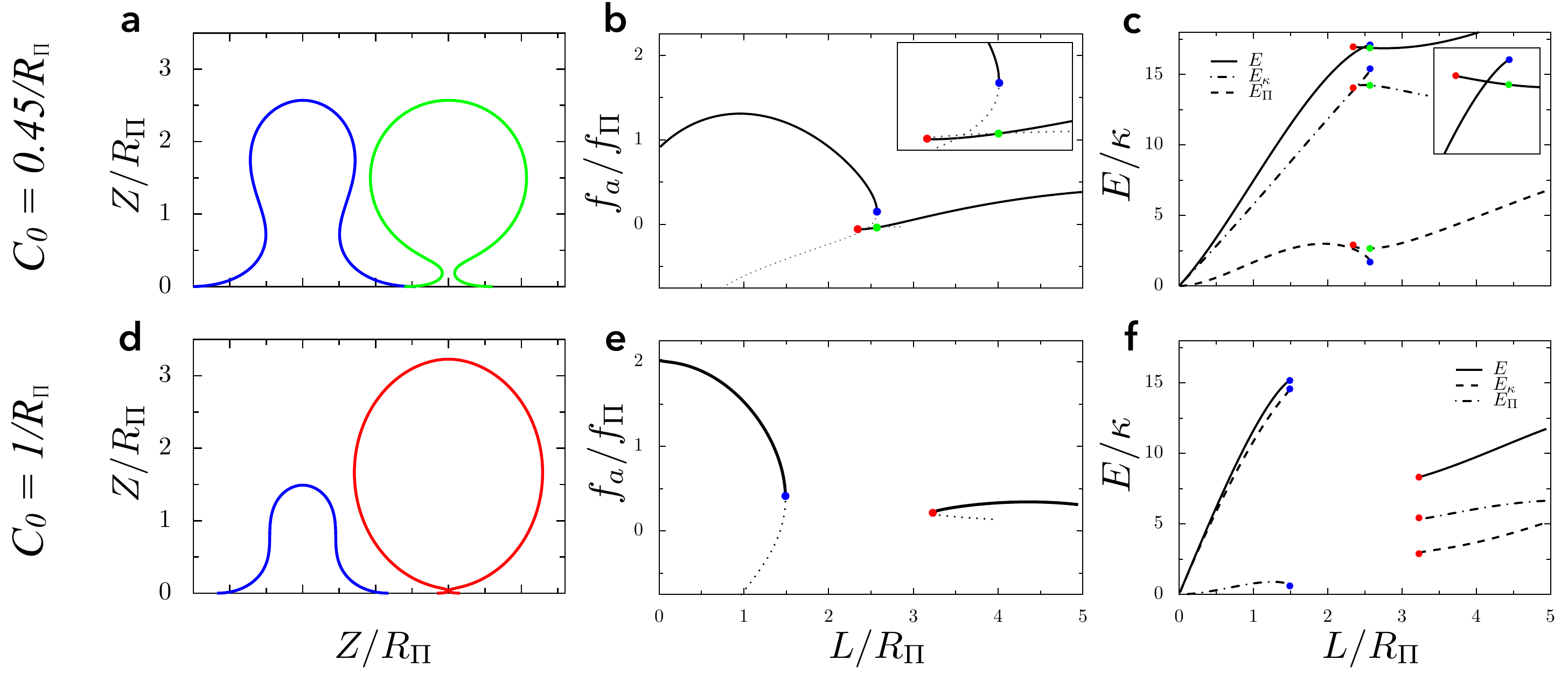}

\caption{\label{transition}
Illustration of the shape instability with $C_0=0.45 / \rpi$ (top) and $C_0=1 / \rpi$ (bottom).
Panels {\bf a, d}: profiles corresponding to the points indicated with the same color on the other panels of this figure.
Panels {\bf b, e}: pulling force (normalized by $f_\Pi$) as a function of invagination length (normalized by $\rpi$).  
With $C_0=0.45 / \rpi$ (top), there is an hysteresis with two branches of solutions: one short and tubular, the other longer and spheroidal. With $C_0=1/\rpi$ (bottom), the two branches do not overlap and no equilibrium shape is found for a range of values of $L$. Dotted lines represent the unstable continuation of the branches. In both cases we took $\sigma=0$. 
Panels {\bf c, f}: Deformation energy as a function of invagination length. The energy associated with membrane rigidity (i.e. bending elasticity) $E_\kappa$, and the one associated with hydrostatic pressure $E_\Pi$ are shown with discontinuous lines. The total deformation energy $E=\mathcal{F}+f_a L$ is plotted with a solid line.
}

\end{adjustwidth}
\end{figure}

\subsection*{Spontaneous curvature leads to neck constriction}

Above a certain height, invagination shapes predicted by theory exhibited a neck, even though there is no constriction force in the model (Fig. \ref{profiles}). 
The only external force is a vertical lifting force applied on the apex of the invagination, and the neck appears as a consequence of the spontaneous curvature (Fig. \ref{profiles}, last profile). 
Higher spontaneous curvatures will produce smaller necks, up to a point where the theory predicts a neck of zero radius, corresponding to a shape instability.

The apparition of a localized neck as a result of global spontaneous curvature has been described theoretically before \cite{bovzivc2014direct}. 
We have more precisely characterized the nature of the shape instability (see supplementary information S.I. 3). 
 We can understand its biological consequences in two limit cases : with a given value of $C_0$ with increasing length $L$, or with a given length $L$ and increasing $C_0$.

\begin{figure}[t]
\centering
\includegraphics[width=11.5cm]{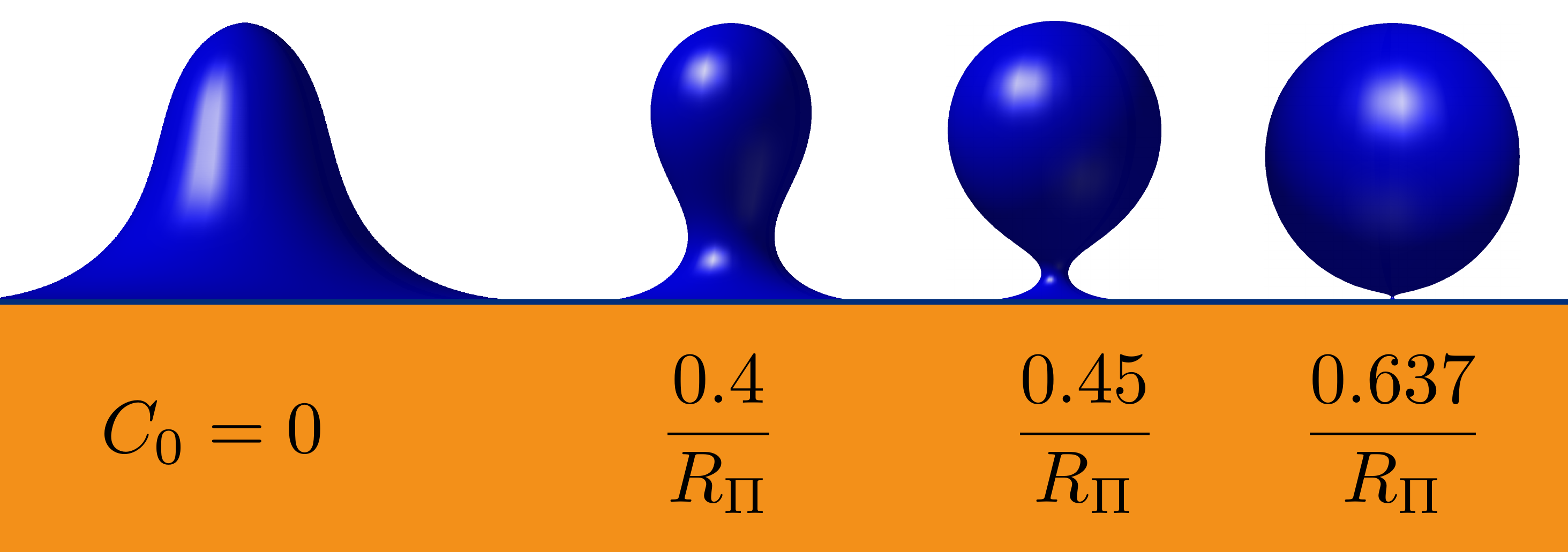}
\caption{\label{increasingC0}
Theoretical membrane profiles for a constant height ($L=2.8 \rpi$) for increasing spontaneous curvature $C_0$, with $\sigma=0$.}
\end{figure}

Firstly, for small values of the spontaneous curvature, the membrane shape evolves smoothly upon increasing the length $L$ of the invagination, and $f_a(L)$ is continuous.
Above a threshold $C_0^*$ however, we can distinguish two branches: one corresponding to short, quasi-tubular invaginations, and another one corresponding to larger spheroidal invaginations (Fig. \ref{transition} {\bf a\,b\,c}, $C_0=0.45/\rpi$). 
The two branches overlap, i.e. for a range of lengths, there are two possible membrane shapes for a given $L$, as illustrated on the example profiles (Fig. \ref{transition} {\bf a}). 
Above a second threshold $C_0^{**}$, the two branches cease to overlap, such that there is a range of $L$ where there is no equilibrium membrane shape (Fig. \ref{transition} {\bf d\,e\,f}, $C_0=1/\rpi$).
The shapes corresponding to the edges of this regions are illustrated in (Fig. \ref{transition} {\bf d}). The transition from the short invagination to the tall spheroidal shape corresponds to a decrease in total membrane deformation energy (Fig. \ref{transition} {\bf f}).  We could compute the values of $C_0^*$ and $C_0^{**}$ as a function of tension (Supplementary Fig. 4).
The same phenomena can be observed for a given length $L$ by increasing $C_0$. 
There is a threshold $C_0^+(L)$ at which the predicted radius is zero, see Fig. \ref{increasingC0}. For larger spontaneous curvatures $C_0 >C_0^+(L)$, no stable membrane shape exist. This is in agreement with the existence of a zero-radius neck at a critical value of $C_0$ \cite{bovzivc2014direct}.  

 This shape instability is not a pearling instability (a shape instability in membrane tubes with tension \cite{bar1994instability}), because it is inhibited by membrane tension (see supplementary information S.I. 3). 
Rather, it stems from the energetic cost of shape defects. Indeed, invaginations cannot be perfectly tubular, and the tip and base of the invagination can be viewed as defects that increase the membrane conformation energy. 
When the invagination takes an almost spherical shape, the tip defect is eliminated, and the cost of the base defect is minimized by making the neck infinitesimal. 
Indeed, we can see that the transition from tubular to spheroidal shapes reduces the total energy as the reduction in bending energy exceeds the increase of pressure-associated energy (Fig. \ref{transition}, {\bf f}).
This membrane shape instability can facilitate membrane scission, a crucial step of endocytosis.

\subsection*{Membrane coating heterogeneity can lead to instability}

Membrane coating is not homogeneous : clathrin, sla2 locates at the tip of invagination, while Rvs167 is usually located at the neck \cite{picco2015visualizing}. Many other proteins can associate to membranes, including the F-Bar protein Bzz1 \cite{kishimoto2011determinants}, epsins such as Ent1 \cite{skruzny2012molecular}, and others  \cite{godlee2013uncertain,kaksonen2005modular}. While we do not know the exact localization, the mechanical properties and the interactions of all these proteins, it is usually believed that the clathrin-coated tip is more rigid than the base of the invagination \cite{walani2015endocytic}. This could also cause a higher spontaneous curvature at the tip, and additional spontaneous curvature could also stem from lipid asymmetry in the membrane leaflets.

We modeled this possible heterogeneity in rigidity as a region with higher rigidity $\kappa$ (the tip), a region with a lower rigidity $\kappa_{min}$ (the base), and a small transition zone (see supplementary information S.I. 1.4). Even in the absence of spontaneous curvature, there is an instability if $\kappa_{min}$ is too small compared to $\kappa$, above a threshold that depends on the surface area having higher rigidity. This instability resembles the curvature instability described above, and also stem from the destabilization of the base of the invagination  (see supplementary information S.I. 1.4).  Overall, a membrane that is more flexible at the base is more prone to shape instability. The results that were derived for a homogeneous membrane, should thus remain qualitatively valid for a heterogeneous membrane as well. 

\subsection*{BAR-domain proteins stabilize the invagination}

BAR-domain proteins are elongated crescent-like objects \cite{mim2012structural}. 
It was shown recently that the number of Rvs167 molecules, a BAR-domain coat protein, increases rapidly in the late stages of the endocytic process \cite{picco2015visualizing}. 
They are thought to induce anisotropic curvature in the membrane.
Following previous work, we assumed that they favor curvature only orthogonally to the symmetry axis ($Z$), i.e. they give a favorite \emph{radius} of curvature $R_0$ \cite{lenz2009membrane}, rather and a favorite total curvature. Noting $\Gamma$ the rigidity of a membrane coated with BAR-domain proteins, the contribution to the energy (eq. \ref{generalff2}) reads :
\begin{eqnarray}
\ff_{\mbox{\scriptsize BAR}} =\iint_S \frac{\Gamma}{2} \left(\frac{1}{R}-\frac{1}{R_0}\right)^2   dS
\label{nrjbar}
\end{eqnarray}

This implicitly assumes that the coverage is uniform on the membrane, whereas in reality Rvs167 proteins are localized at the neck of the invagination \cite{picco2015visualizing}.
This simplification was necessary however to derive the membrane shape equations (see supplementary information S.I. 1.2) and sufficient to understand in essence how anisotropic curvature can affect the invagination. 
It allowed us predict the membrane shape in the presence of both isotropic curvature effectors (with parameters $C_0, \; \kappa$) and anisotropic curvature effectors (with parameters $R_0, \; \Gamma$).

\begin{figure}[t]
\centering
\includegraphics[width=11.5cm]{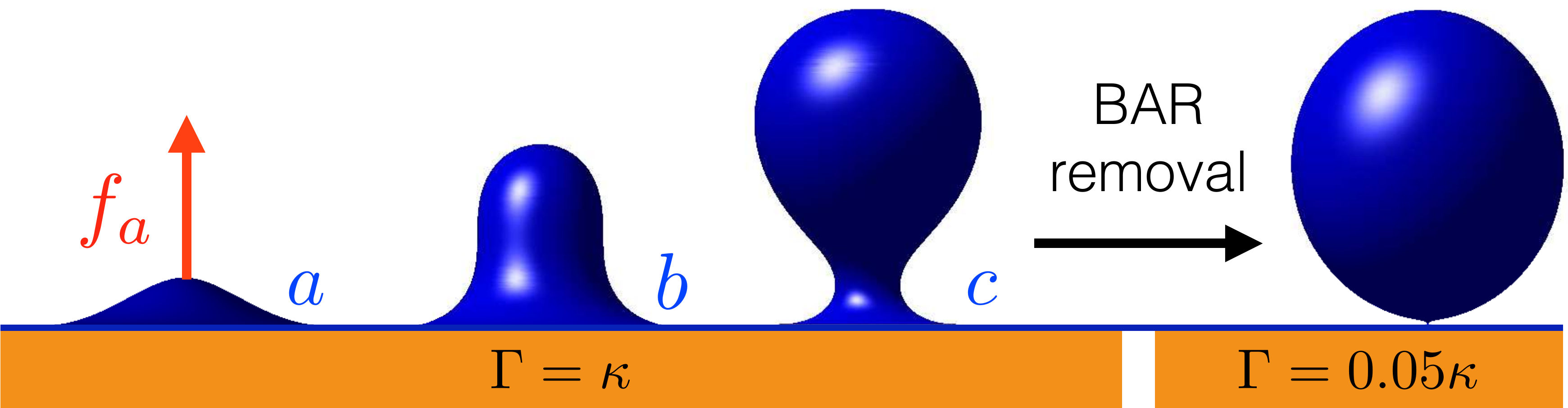} \\
\includegraphics[width=11.5cm]{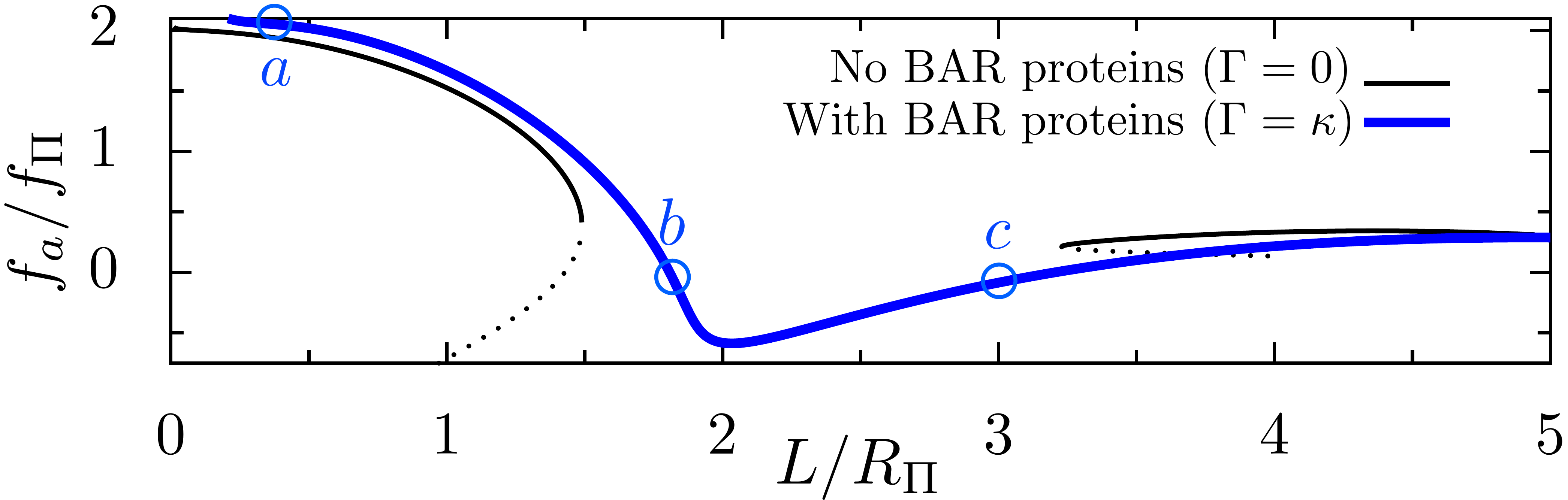}
\vskip 0.2cm
\caption{\label{force_bardomains}
{\bf Top}: Theoretical membrane profiles for a membrane coated both with isotropic curvature effectors (such as clathrin) with $C_0=1/\rpi$ and anisotropic membrane effectors (such as bar domain proteins) with $R_0=\rpi$. The first three shapes correspond to three steps of invagination process with $\Gamma=\kappa$. The last shape was obtained with $\Gamma \rightarrow 0$, {\it i.e.} by removing the BAR-domain proteins. The removal leads to an infinitesimal neck that promote scission.
{\bf Bottom}: Required force as a function of invagination height, with and without anisotropic spontaneous curvature ($\Gamma=0$ or $\kappa$), using $\sigma=0$, $C_0=1/\rpi$ and $R_0=\rpi$.
In the presence of anisotropic curvature induced by BAR-domain proteins, there is no shape instability (the blue curve is continuous).
Comparing with the black curve shows that the shape instability was suppressed.
}
\end{figure}

We found that anisotropic curvature inhibit the membrane shape instability. The membrane is therefore stabilized and longer invaginations can grow continuously even for high values of $C_0$ (Fig. \ref{force_bardomains}, bottom, where we used $R_0=\rpi$ for simplicity). 
This is in agreement with {\it in vivo} observation that invaginations without Rvs167 cannot grow longer than about $60\; nm$.
As a corollary, removing BAR-domain proteins is a possible mean of triggering membrane scission.
In our theory, this corresponds to $\Gamma \rightarrow 0$, which indeed destabilizes the membrane (Fig. \ref{force_bardomains}, top). 
This possibility is supported by the recent observation that membrane scission is synchronous with the disappearance of Rvs167  \cite{picco2015visualizing}.

\section*{Discussion}


Using a general model for membrane mechanics, we could accurately fit experimental profiles shorter than $80 nm$, even though we assumed the membrane to be homogeneous, i.e. with constant rigidity and spontaneous curvature over the surface of the deformed membrane. In combination with membrane rigidity, we can expect pressure to be the dominant factor opposing membrane invagination during yeast endocytosis, while membrane tension should be negligible. 
This statement is derived from the dimension of the invagination, and from the scale of pressure determined experimentally, and is thus independent of the details of the model.
We estimated the force required to pull the invagination based on the value of turgor pressure, and on the value of the rigidity that was determined by fitting the experimental curves. 
While the exact value of the force also depends on other parameters and in particular on the height of the invagination (see Fig. \ref{illus_force_length}), its scale is primarily determined by pressure and the width of the invagination.
For the measured range of pressure $\Pi \sim 0.2\text{---}1  \MPa$, the force scale is $f_\Pi \sim 1000\text{---}5000\; pN$.
This is significantly larger than previously estimated ($1 \mbox{ --- } 1000\; pN$ \cite{liu2006endocytic,carlsson2014force,liu2009mechanochemistry}). 
The corresponding range of value for the rigidity is $\kappa \sim 400\text{---}2000\; k_B T$.
This is much stiffer than a pure lipid bilayer, because the membrane is heavily coated at the endocytic site.
It was suggested that phase boundaries could play a role by generating a line tension, favorable to membrane budding. 
However the typical line tension (of the order of $\sim 0.4\; pN$ \cite{tian2007line}) is much smaller than $f_\Pi$, and such phenomenon can therefore be discarded. 


The fact that the scale of the invagination is determined by the ratio of rigidity and pressure ($\rpi = \sqrt[3]{\kappa/2\Pi}$) suggests a possible tradeoff.
In the absence of any reinforcement, the naked membrane could only make small invaginations with a width of $3\; nm$. 
The coat enlarges the invagination by a factor $\sim 10$ by increasing the local rigidity, but at the same time also increases the required pulling force. 
The mechanical properties of the coat thus likely represent an optimal tradeoff between increasing the radius of the invagination (making individual endocytic events more productive) and limiting the driving force required from the actin machinery.
The ratio $\kappa/\Pi$ is likely to have been adjusted in the course of evolution, and is not expected to be conserved across species.

 In general, coat protein can induce a negative tension that can favor tubulation \cite{lipowsky2013spontaneous}, because of their adhesion energy with the membrane. This adhesion energy has been estimated to $\omega \sim 10^{-4} N/m$ in the case of clathrin \cite{saleem2015balance}, while other coat protein may have adhesion energies of $\omega \sim 10^{-3} N/m$ \cite{ewers:2009}.  Using the largest of the two values, we find a pulling force $\pi \omega R \sim 30 pN$, which is insufficient to drive the invagination against the turgor pressure. Moreover, we found that spontaneous curvature actually increases the initial pulling force in these conditions, and consequently coat proteins do not help to initially lift the membrane. This counter-intuitive result is in agreement with observations that clathrin is not necessary to initiate curvature during endocytosis in yeast \cite{kukulski2012plasma}. This result is specific to the situation where a membrane subjected to a large pressure has to be pulled away from its supporting wall. In this configuration, the invagination must have regions with positive (at the tip) and negative (at the rim) curvature, and the energetic cost of the latter are significant in the presence of positive spontaneous curvature.

The good quality of the fits (Fig. \ref{profiles}) indicates that the experimental invagination profiles are consistent with a pulling point force, directed away from the cell wall. 
In the cell, the forces generated in the actin network are probably transferred to the tip of the invagination over an finite area corresponding to the sla2 proteins. 
In the model, the point force is effectively propagated over the tip of the invagination, due to the inferred rigidity of the coated membrane.
As long as the coat structure remains sufficiently rigid, these two situations should be equivalent.

We find that under physiological conditions of pressure, direct constriction by actin in a mechanism similar to cytokinesis is not necessary to explain the shape of the invagination, or specifically the formation of a neck. 
This result is similar to what was reported in the absence of turgor \cite{bovzivc2014direct}, and stems from having an isotropic spontaneous curvature that is everywhere the same on the membrane.

 With isotropic spontaneous curvature, the membrane can have a shape instability (Figs. 4 and 5). A more complicated model that included some level of inhomogeneity (S.I. 1.4), showed that this shape instability is facilitated if the coated membrane is less rigid at the base than at the tip of the invagination (S.I. 3.2).
We thus expect the shape instability to be promoted by the inhomogeneity of the system, in particular the fact that the membrane tip is covered with clathrin while the base most likely is not. 
Qualitatively, pressure is pushing on the neck, thus constricting it, while membrane tension is pulling on it.
Thus the shape instability is promoted by pressure and inhibited by membrane tension (S.I. 3.1).

Crescent-shaped curvature effectors such as Rvs167 were modelled as inducing anisotropic curvature. 
The analysis showed that their presence stabilizes the invagination, suggesting that Rvs167 removal can cause membrane destabilization and scission.  It is known that membrane fission is promoted by amphipathic helix insertion and inhibited by crescent BAR domains \cite{boucrot2012membrane}, and a theoretical argument was made comparing idealized tubes to spherical vesicles. Here we show that crescent-shaped proteins also stabilize intermediate shapes of endocytosis, in what appears to be a very generic phenomenon.

It was reported that Rvs167-deprived cells have shorter invaginations \cite{kukulski2012plasma}. 
A possible explanation is that BAR domain proteins stabilize invaginations, as in our theory.
An alternative possibility is that Rvs167 would be needed to lower the pulling force required at later stages of endocytosis. 
However, our calculations indicate that the highest forces are required during the initial stages of endocytosis (Fig. \ref{illus_force_length}). 
Indeed experimentally, the growth speed of the later stages seems independent from the pressure \cite{basu2014role}, which is the main force opposing endocytosis. 
This indicates that stabilization of the shape by BAR domains, rather than reduction of force, is the most likely explanation of the observed shortening of invaginations in Rvs167 mutants. 

The homogeneous theory could not fit large membrane deformations ($>80nm$).
Indeed, Rvs167 appears at the neck for large invaginations \cite{picco2015visualizing}, and probably induce inhomogeneous properties near the base. 
This effect was not included in the inhomogeneous model described in (S.I. 1.4), because introducing several different regions makes it technically intractable.

Endocytosis is ubiquitous in nature and in other model systems such as animal cells, it appears to involve a similar set of molecules. The physical conditions can however vary greatly in different cells, and this should be reflected in some of the requirements that have constrained the evolution of the endocytic machinery. In the absence of significant hydrostatic pressure, membrane coating is sufficient to generate invaginations \cite{ewers:2009}, while dissipative processes involving actin and/or dynamin machinery appear necessary to membrane severing \cite{grassart2014actin,ferguson2012dynamin}. 
For the yeast system that we examined, pulling the membrane away from the cell wall seems sufficient to induce a complete budding event. The next crucial step is to decipher the mechanism by which the actin cytoskeleton is able to exert the required force. 
Precise quantitative information is available to do so \cite{berro2014, picco2015visualizing} and theoretical work is under way \cite{carlsson2014force}. 
Ultimately, this will offer a unified model incorporating both actin cytoskeletal dynamics and membrane mechanics, based on the experimental observations. 

\section*{Supplementary information}

\begin{itemize}
\item {\bf S.I. 1 Membrane shape} 
\item {\bf S.I. 2 Forces}
\item {\bf S.I. 3 Shape instability}

\end{itemize}

\section*{Author contributions}
S.D. and F.J.N. designed research; S.D. performed research; Both wrote the paper together.

\section*{Acknowledgments}
We thank Wanda Kukulski, Andrea Picco and Martin Schorb for stimulating discussions and technical assistance; Pierre Sens, John Briggs, Marko Kaksonen, Andrea Picco and Ori Avinoam 	for their critical reading of the manuscript. We thank EMBL IT services for their support.

\newpage
\section{S.I. : Membrane shape}
\subsection{Differential equations}
\begin{figure}[t]
 \includegraphics[width=5cm]{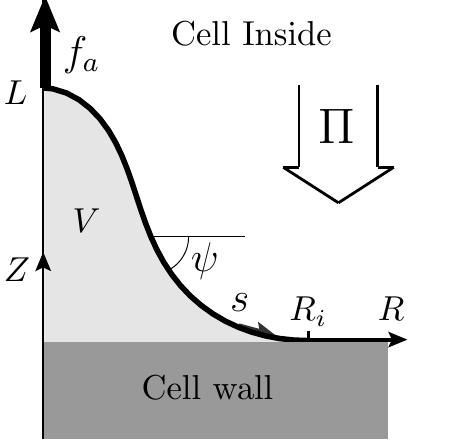}
\caption{\label{illus_param}
Diagram of the endocytic process and its parametrization. We assumed rotational symmetry around the $Z$ axis, normal to the cell wall. The shape is parametrized by the curvilinear abscissa $s$, the distance $R$ from the symmetry axis, and the angle $\psi$. $L$ is the maximal height of the invagination, $R_i$ is the distance beyond which the membrane is in contact with the cell wall, and $\Pi$ is the difference of hydrostatic pressure across the plasma membrane.}
\end{figure}

The membrane shape equations correspond to the minimum of the membrane energy (main text, equation 1). To derive these equations, we renormalize energies by $\kappa$ and distances by $\rpi$, without loss of generality. The typical scale of forces will be $f_\Pi = 2 \pi \kappa / R_\Pi$.
Following earlier work  \cite{julicher1994shape}, we assume rotational symmetry around the vertical axis, which is defined as the direction normal to the cell wall, orientated towards the cell inside. Let $Z$ be the distance from the cell wall along this axis, $R$ the distance from the symmetry axis, $s$ the arclength along the membrane, and $\psi$ the angle of the membrane with the horizontal axis (see Fig. \ref{illus_param}, left), all being dimensionless quantities. Additionally, we define  $\bar{\sigma}={\rpi}^2 \sigma / \kappa$, $f=f_a / f_\Pi$, $\bar{C_0}=\rpi C_0$ and $\bar{\Pi}=1/2$. To impose the geometric relations  $\partial R/\partial s = \dot{R}=\cos{(\psi)}$ and $\dot{Z}=-\sin{(\psi)}$, we use two Lagrange multipliers $\nu$ and $\eta$.  
The renormalized energy $\ff'$ finally reads :
\begin{equation}
\ff' = 2 \pi \int_S \mathcal{L}(R,\psi,\dot{\psi}) ds \, , \label{generalffSI}
\end{equation}
with $\mathcal{L}$ the renormalized energy density:
\begin{equation}
\mathcal{L} = \frac{R}{2} \left( \dot{\psi} + \frac{\sin{\psi}}{R} - \bar{C_0} \right)^2 + \bar{\sigma} R  + \frac{\bar{\Pi} R^2}{2}\sin{\psi} -f \sin{\psi}  + \nu (\dot{R} - \cos \psi) + \eta (\dot{Z} + \sin \psi) \, ,\label{lagrangeff}
\end{equation}
where the dot indicate the derivative with respect to the arc length $s$.
From these, we derive the Euler-Lagrange equations \cite{julicher1994shape} :
\begin{eqnarray}
\ddot{\psi} = \frac{\cos \psi }{R} \left( \frac{\sin \psi}{R} - \dot{\psi} + \eta-f \right)
+ \frac{\bar{\Pi}}{2} R \cos \psi + \frac{\nu}{R} \sin \psi \, ,\label{EL1}\\
\dot{\nu} = \frac{1}{2} (\dot{\psi}-\bar{C_0})^2 - \frac{\sin^2 \psi}{2 R^2} + \bar{\sigma} + \bar{\Pi} R \sin \psi\, , \label{EL2}  \\ 
\dot{\eta} = 0 \, ,  \label{EL3} 
\end{eqnarray}
which are valid between $R=0$ (the center of the invagination) and $R=R_i$, after which the membrane is in contact with the cell wall. 
 Since the energy involves $\dot{\psi}$, $\psi$ has to be continuous and hence $\psi=0$ at $R=R_i$, whereas $\dot{\psi}$ can be discontinuous. Canceling the boundary term $[\nu R_i]$ imposes $\nu(0)=0$ (as $R_i$ is a free parameter) and canceling the Hamiltonian \cite{julicher1994shape} imposes $\dot{\psi}(0)=-\sqrt{\bar{C_0}^2 + 2 \bar{\sigma} }$.  For simplicity, we assumed $\bar{C_0}$ to be constant over the whole membrane surface, an approximation discussed later.

While assuming axisymmetry may seem a strong approximation, minimizing Helfrich's Hamiltonian will produce axisymmetric shapes, and a more general theory would yield the same result, unless the force $f_a$ is not orthogonal to the cell wall, or if the membrane properties are severely anisotropic \cite{dommersnes1999n}. Experimentally, we see membrane profiles that are not perfectly axisymmetric \cite{kukulski2012plasma}, which is expected considering the small size of the actin machinery, where possibly as few as 10 filaments are polymerising at any time \cite{berro2014}.

\subsection{Membrane with anisotropic coat}
To describe a membrane coated with anisotropic coat proteins such as Rvs167, we added a term favoring a radius $R_0$ to the energy density \cite{lenz2009membrane} :
\begin{eqnarray}
\mathcal{L}'= \mathcal{L}+\frac{\bar{\Gamma}}{2} R \left( \frac{1}{R} - \frac{1}{\bar{R_0}} \right)^2 \, ,
\end{eqnarray}
in which $\mathcal{L}'$ is the new renormalized energy density, $\bar{\Gamma}=\Gamma/\kappa$ and $\bar{R_0}=R_0/R_\Pi$.

The differential equation for $\ddot{\psi}$ is unchanged (equation \ref{EL1}) but a term $\frac{1}{2}\bar{\Gamma}(1/R-1/\bar{R_0})^2$ is added to $\dot{\nu}$ (equation \ref{EL2}). Additionally, the boundary value of $\dot{\psi}$ is now :
\begin{eqnarray}
\dot{\psi}(0)= - \sqrt{\bar{C_0}^2 + 2 \bar{\sigma} + \bar{\Gamma} \frac{(R-\bar{R_0})^2}{R^2 \bar{R_0}^2}}
\end{eqnarray}

\subsection{Numerical solutions}
To find the membrane shape, we integrated equations \ref{EL1}-\ref{EL3} from $R=R_i$ down to $R \rightarrow 0$, i.e. taking a negative $ds$. Because of the divergence at $R=0$, in practice we integrated from $R_i$ to a non-zero cutoff $\epsilon= 0.004$, that was chosen to be smaller than a protein. We used a shooting method \cite{press2007numerical} to find the values of $R_i$ and $f$ such that the membrane reaches $Z=L$ and $\psi=0$ at $R=\epsilon$. Because of the strong non-linearity in the equations, most initial values of $R_i$ and $f$ will lead to a diverging shape. However, by scanning randomly for initial values of $R_i$ and $f$ for a given $L$, we could find parameters that gave a physical solution to the equations.

 To compute membrane shape for a range of $L$ for a given value of $\bar{C_0}$ and $\bar{\sigma}$, we started from a small value of $L$ and determined the best $\{R_i,f\}$ by a random search coupled to a shooting method. Then, we looked for the solution for $L+\delta L$ using the shooting method starting from the same $R_i$ and $f$. When $\delta L$ was small (typically $10^{-6} \mbox{---} 10^{-3} R_\Pi$), the shooting method usually converged, unless a bifurcation was reached. 
 To find solutions for other parameters $\bar{C_0}$ and $\bar{\sigma}$, we either started another random parameter search coupled to the shooting method, or started from a known solution, and slowly changed the $\bar{C_0}$ and/or $\bar{\sigma}$ (again with small steps).

\subsection{Heterogeneous membrane}

To model the existence of membrane heterogeneities, namely the clathrin-rich tip of the invagination, we introduced a non-constant rigidity $\kappa \mapsto \kappa \,  \alpha(s)$.
Assuming $\alpha(s)$ to be derivable, this changes the differential equations for $\ddot{\psi}$ :
\begin{eqnarray}
\ddot{\psi} = \frac{\cos \psi }{R} \left( \frac{\sin \psi}{R} - \dot{\psi} + \frac{\eta-f}{\alpha} \right)
+ \frac{\bar{\Pi}}{2\alpha} R \cos \psi
+ \frac{\nu}{\alpha R} \sin \psi 
- \frac{\dot{\alpha}}{\alpha} \left( \dot{\psi} + \frac{\sin \psi}{R} - C_0 \right) \label{EL1_bis} \\
\dot{\nu} = \frac{\alpha}{2} (\dot{\psi}-\bar{C_0})^2 - \alpha \frac{\sin^2 \psi}{2 R^2} + \bar{\sigma} + \bar{\Pi} R \sin \psi  \label{EL2_bis}  \\ 
\dot{\eta} = 0 \, . \label{EL3_bis} 
\end{eqnarray}

For $\alpha(s)$ we used a function varying smoothly between $1$ and $\alpha_{min}$, over a width $\Delta s$ :
\begin{eqnarray}
\alpha (s) =1 \qquad  \quad  \text{if } s \le s_1  \\
\alpha (s) = 1 + (\alpha_{min} - 1) \left( 1- e^{-\frac{(s_1 - s)^2}{2 \Delta s^2} } \right) \qquad \text{if }  s>s_1 
\end{eqnarray}
Here, $s_1$ is the position of the step along the arclength (with $s=0$ at $R=R_i$ and $s=s_{tot}$ at $R=\epsilon$). We determined $s_1$ numerically to obtain that the surface area defined by $s < s_1$ is the tip surface $S_{tip}$.

\subsection{Fitting of experimental data}

The 2D experimental profiles derived from electron microscopy imaging \cite{kukulski2012plasma} are available online. They needed to be projected to an axisymmetric profile, since the theory could only consider axisymmetric shapes. For a given profile, we find the two points A,B where the membrane intersects an horizontal axis, at each height from the base. We then extract the two angles of the membrane with respect to the vertical axis at points A and B, and we define the local orientation $t$ as the average of these angles.  The radius of an axisymmetric profile at this height is then $[AB] \cos{(t)}/2$.

\begin{figure}[t]   
\begin{adjustwidth}{-2.25in}{0in}
\centering
\includegraphics[width=7cm]{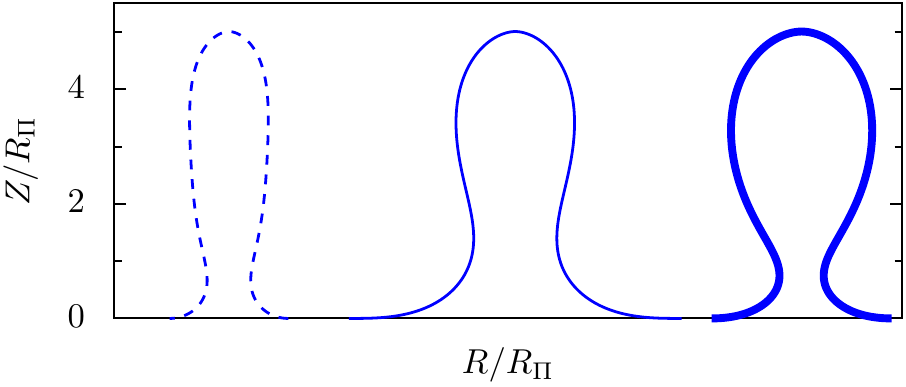} \\
\includegraphics[width=7.2cm]{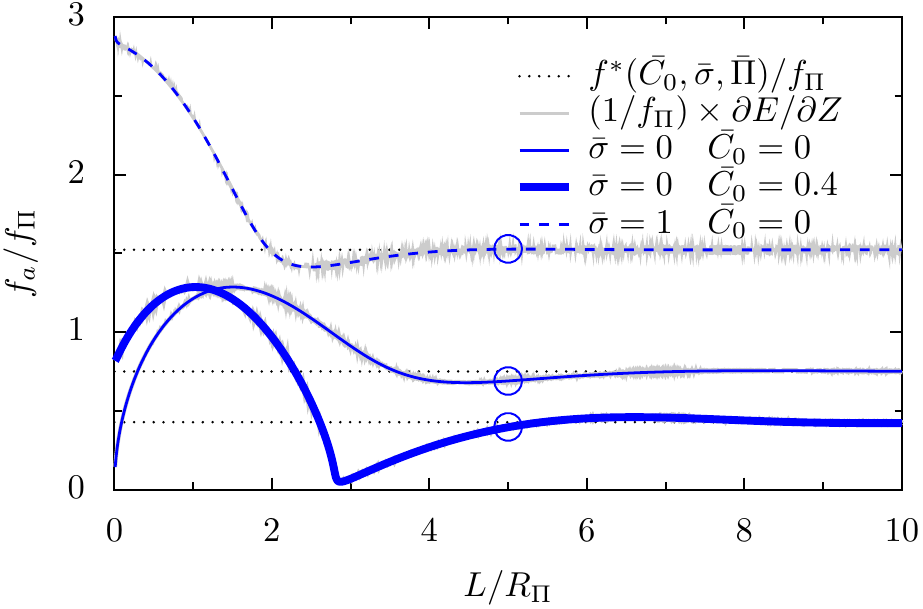} 
\caption{\label{illus_force_shape}
Top : three membrane shape profiles. Thin line: $\bar{C_0}=0,\bar{\sigma}=0$, dashed line: $\bar{C_0},\bar{\sigma}=1.0$, and thick solid line: $\bar{C_0},\bar{\sigma}=0$. 
Bottom : force versus height of the invagination, for the three aforementioned parameter sets. The gray and noisy lines represent the numerical calculation of $\partial E / \partial L$, showing that our results are thermodynamically coherent. Dotted lines indicate the analytical predictions of the plateau force. 
 }
 \end{adjustwidth}
\end{figure}

In the fitting procedure, it is important to note that $\rpi$ act as a scaling parameter, while $\bar{\sigma}$ and $\bar{C_0}$ control the shape of the invagination. The renormalized force $f$ can be seen as a Lagrange multiplier controlling the height of the invagination. Therefore, we compared experimental shapes to theoretical profiles with a large range for $\bar{\sigma}$ and $\bar{C_0}$, and for each profile, we tried values of $R_\Pi$ from $5nm$ to $35nm$.  For a given set $\{ \bar{C_0},\bar{\sigma},R_\Pi \}$, only one value of $f_a$ produces the correct invagination height. 

The error between a theoretical profile and an experimental shape was defined as :
\begin{eqnarray}
\text{Error}=\sum_i \frac{(R_i^{th}-R_i^{exp})^2}{(R_i^{exp})^2} 
\end{eqnarray}
Where the $R_i$ are the radii of profile sections, for height between $0$ and $L$, where $L$ is the length of the invagination.

A range of parameters $\{ \bar{C_0},\bar{\sigma},R_\Pi \}$, rather than a discrete set, can be used to fit each profile, while keeping the error below $5\%$. We found that most profiles could be fitted with $R_\Pi \sim 15-25 nm$, $\bar{\sigma} \sim 0$, and $\bar{C_0} \sim 0 \text{--} 0.5$. 
The accuracy of the fit is undermined by the facts that we did not fit the original profiles but an axisymmetric projection. Moreover the solution space is degenerate, and bias are possible in the experimental methods.

\section{S.I. : Forces}
\subsection{Force as the derivative of the energy}

In our physical theory, the force $f_a$ pulling the invagination should satisfy the thermodynamic definition of a force, i.e. $f_a =-\partial E / \partial Z$, in which $E=\mathcal{F}+f_a L$ is the total internal energy. Hence $f$ can also be seen as a Lagrange multiplier controlling the height of the invagination. We have several ways to test the validity of our force computation. First, we can compute analytically the force required to pull long invaginations ($L \gg  R_\Pi$), that are mostly tubular \cite{derenyi2002formation}. In this configuration, the total energy of the system is dominated by the energy of the tubular portion, and the radius of the tube $R^*$ can be found by minimizing the energy per unit length of tube:
\begin{eqnarray} 
E_{tube} = 2 \pi R  \left[ \frac{\kappa}{2} \left( \frac{1}{R} - C_0 \right)^2 + \sigma \right] + \Pi \, \pi R^2 \label{etube}\\
\partial_R E_{tube} \bigg|_{R^*} = 0 \, . \label{tuberad}
\end{eqnarray}
The numerically computed plateau value for the force matches this analytic value $f^*= E_{tube} (R^*)$ perfectly (figure \ref{illus_force_shape}, bottom).

Second, to confirm that the force computation is exact for any $L$, we verified that the input value for $f_a$, found by the shooting method, corresponded to the thermodynamical definition of the force $\partial_L E$. 
The numerical estimation of $\partial_L E/ f_\Pi$ is noisy (due to small error of on the height $L$ and on the angle $\psi$ at the tip), but indeed $f$ always matches $\partial_L E / f_\Pi$ (figure \ref{illus_force_shape}, right, underlying grey lines).

\begin{figure}[t]   
\centering
  \includegraphics[width=8.0cm]{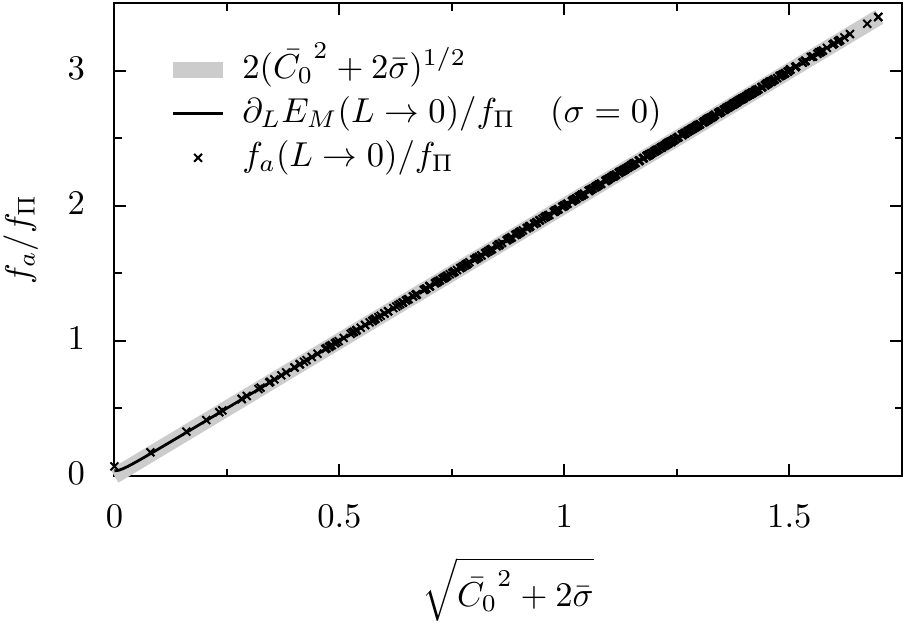} 
\caption{\label{illus_force_init}
Initiation force $f_0$ as a function of physical parameters $\bar{C_0}$ and $\bar{\sigma}$. Black line : numerical solution in the Monge approximation. Crosses : numerical solution of the general shape equations. Grey line : fit by $f_0 = 2 f_\Pi \sqrt{\bar{C_0}^2 + 2 \bar{\sigma}}$. }
\end{figure}

\subsection{Initiation force}

At the early stages of endocytosis, before the appearance of a neck, we can describe the membrane in the Monge form using $Z(R)$ and $Z'=\partial Z / \partial R$ \cite{nelson1987fluctuations}. For early invaginations, we can further take the small deformation limit $Z'\rightarrow 0$, leading to a simpler formulation of the energy:

\begin{eqnarray} 
E_M \sim   2 \pi \kappa \int_0^{R_i} \left[ (\Delta Z - \bar{C_0})^2 2 \bar{\sigma} {Z'}^2 + 2 \bar{\Pi} Z \right] R d R \, . 
\end{eqnarray}

Taking the limit $\bar{\sigma} \rightarrow 0$, equations \ref{EL1},\ref{EL2} become, for an invagination of renormalized length $\bar{L}$ :
\begin{eqnarray}
Z(R) = -\frac{1}{128} R^2 (R^2 + 32 a ) + \bar{L} +b \, R^2 \log(R) \, .  \label{mongeshapeq}
\end{eqnarray}

\begin{figure}[t]
\begin{adjustwidth}{-2.25in}{0in}
 \includegraphics[width=8cm]{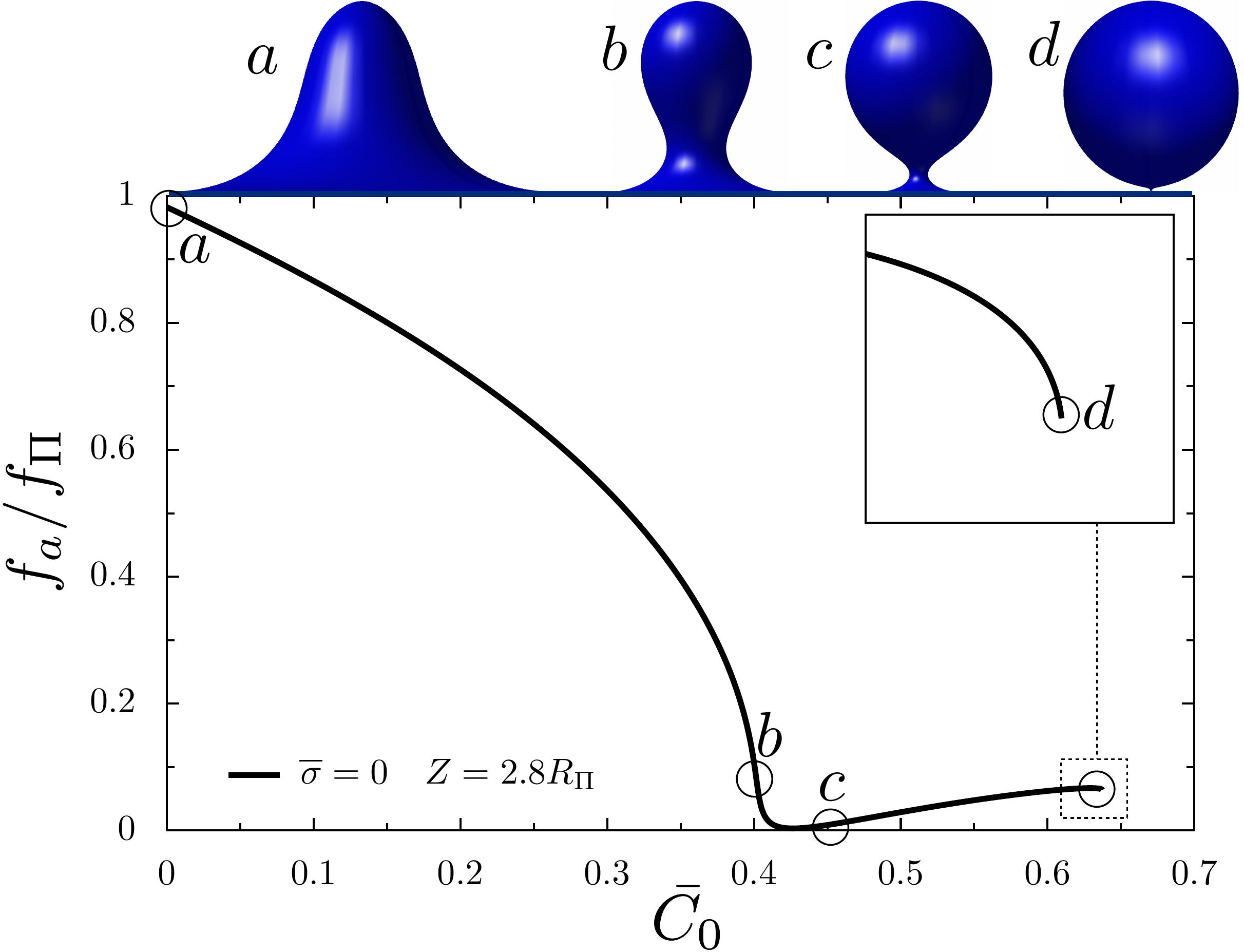} 
 \includegraphics[width=9cm]{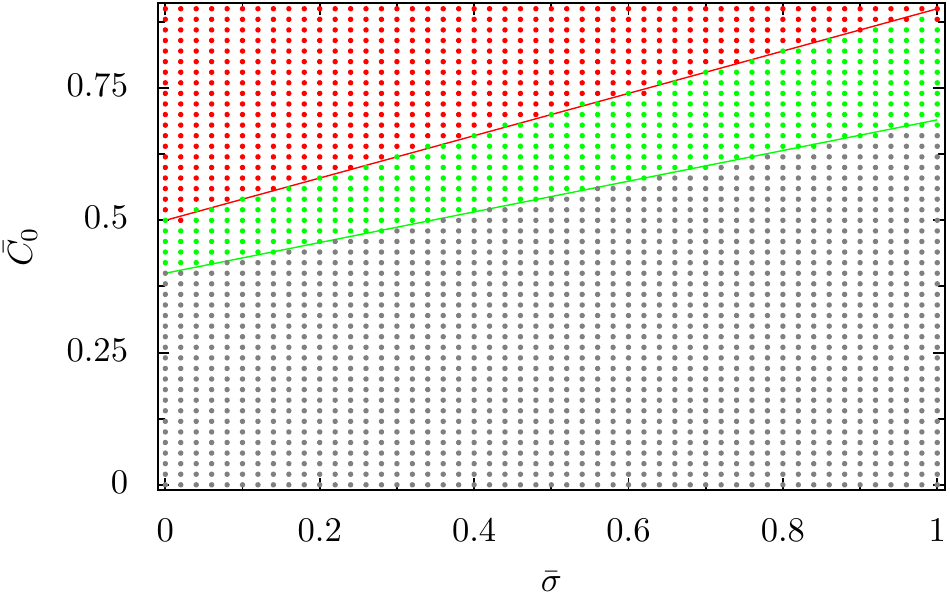} 
\caption{\label{illus_scission}
Left : Force as a function of spontaneous curvature for $L=2.8 \rpi$ and $\bar{\sigma}=0$; the close-up shows that the branch ends at a finite $\bar{C_0}$. Right: Phase diagram of the budding instability as a function of spontaneous curvature $\bar{C_0}$ and tension $\bar{\sigma}$. Grey: no instability, green: instability with hysteresis, red: instability with discontinuity (lack of solutions for a range of $L$). Solid lines are visual guidelines showing that the critical value of $\bar{C_0}$ depends linearly on $\bar{\sigma}$.}
\end{adjustwidth}

\end{figure}
The parameters $\{a,b,R_i\}$ should satisfy the boundary conditions :
\begin{eqnarray}
 Z(R_i)=Z'(R_i)=0 \, ,\\
Z''(R_i)=-\sqrt{\bar{C_0}^2 + 2 \bar{\sigma} } \, .
\end{eqnarray}
For any given $\bar{L}$, we can then compute numerically $Z(R)$, from which we deduce $E_M(\bar{L})$. We can then compute the force $\partial E_M / \partial \bar{L}$ numerically. We could compare the approximate result from the Monge representation (in the absence of tension), the numerical result from the general membrane shape, and we find that they are well fitted by the formula :
\begin{equation}
f_0 = 2 f_\Pi \sqrt{\bar{C_0}^2 + 2 \bar{\sigma}}
\end{equation}

%


\section{S.I : Shape instability}
\subsection{Curvature-induced shape instability}
It is interesting to compare the shape instability that we found with the pearling instabilities (see \cite{tsafrir2001pearling} and references therein), whereby (local) spontaneous curvature competes with membrane tension \cite{bar1994instability,derenyi2007membrane}, usually under constraints of volume conservation \cite{tsafrir2001pearling}. Comparing the energy of a tube (equation \ref{lagrangeff} with $\psi=\pi/2$ and $\dot{\psi}=0$) to that of a string of spheres does not yield any instability in our case. Instead, upon increasing $L$ at the transition, the bending energy decreases more than the pressure energy increases, and this transition occurs at heights where the membrane is far from tubular. This transition is therefore caused by the cost of bending the membrane at the tip and tail of the invagination, and these defects cost more energy in the quasi-tubular shape than in the spheroid shape, where the neck is a defect that can become infinitesimally small. The phase diagram (Fig. \ref{illus_scission}, right), shows that membrane tension actually inhibits this transition, which is due to membrane tubes of minimum energy having less membrane area than spheroids. The phase diagram also exhibit a region in which there is no stable membrane shape for a range of invagination heights, in contrast with the pearling instability. Despite the complexity of the equations, the phase boundaries exhibit a linear behavior for the critical value of $\bar{C_0}$ as a function of $\bar{\sigma}$ (Fig. \ref{illus_scission}, right).

\subsection{Rigidity-induced shape instability}

Strikingly, there is a shape instability even in the absence of spontaneous curvature if the membrane is heterogeneous, with a basal membrane of lesser rigidity $\kappa_{min}=\kappa \alpha_{min}$. This instability appears if $\kappa_{min}/  \kappa$ is smaller than a certain threshold, which depends on the surface area with higher rigidity. For example, with tip surface of  $2 \pi R_\Pi^2$, we found a threshold of $\kappa_{min}^* \sim 0.25 \kappa$. 

\begin{figure}[t]

\begin{adjustwidth}{-0in}{0in}
\centering
\includegraphics[width=12cm]{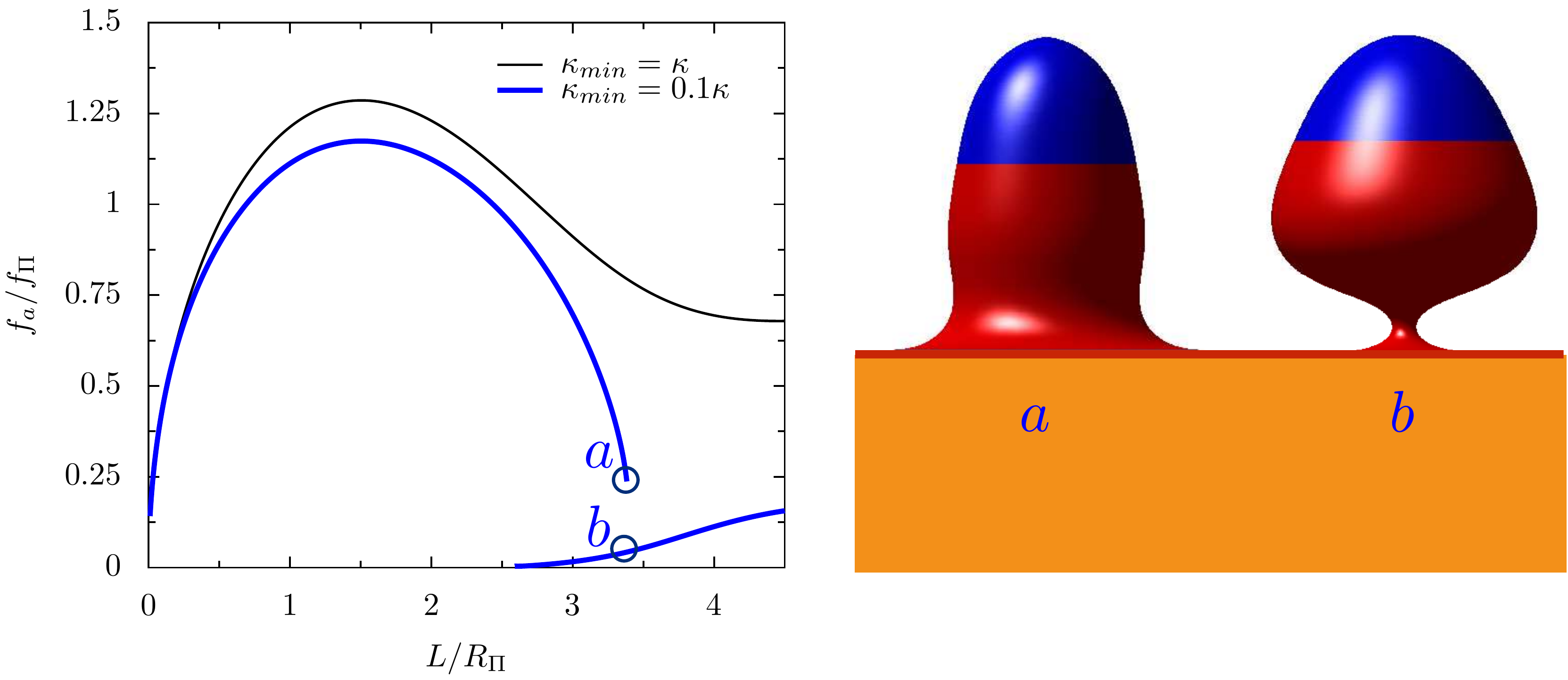} 
\caption{\label{illus_hetero} 
Left : pulling force (normalized by $f_\Pi = 2 \pi \kappa / \rpi$) as a function of invagination length (normalized by $R_\Pi$). The blue line corresponds to a heterogeneous membrane (with a rigid tip of surface $S_{tip}=2 \pi R_\Pi^2$), while the black line shows the behavior for a homogeneous membrane, both in the absence of spontaneous curvature. Right : membrane profiles for the two solution branches. The blue surfaces indicate the rigid tips of the invaginations, while the red surfaces correspond to the more flexible regions. 
}
\end{adjustwidth}
\end{figure}

We then used $\kappa_{min}=0.1 \kappa$ (representative of the ratio of clathrin rigidity to membrane rigidity). Though the membrane shapes are greatly influenced by membrane heterogeneity (Fig. \ref{illus_hetero}, right), their behavior as well as that of the force-distance curve (Fig. \ref{illus_hetero}, left) is very similar to the curvature-induced instability, and it appears that both instabilities are of the same nature.

\bibliographystyle{unsrt}
\bibliography{BibTeX_ref_pio}

\begin{thebibliography}{10}

\bibitem{mukherjee1997endocytosis}
Sushmita Mukherjee, Richik~N Ghosh, and Frederick~R Maxfield.
\newblock Endocytosis.
\newblock {\em Physiological reviews}, 77(3):759--803, 1997.

\bibitem{mcmahon2011molecular}
Harvey~T McMahon and Emmanuel Boucrot.
\newblock Molecular mechanism and physiological functions of clathrin-mediated
  endocytosis.
\newblock {\em Nature reviews Molecular cell biology}, 12(8):517--533, 2011.

\bibitem{sirotkin2010quantitative}
Vladimir Sirotkin, Julien Berro, Keely Macmillan, Lindsey Zhao, and Thomas~D
  Pollard.
\newblock Quantitative analysis of the mechanism of endocytic actin patch
  assembly and disassembly in fission yeast.
\newblock {\em Molecular biology of the cell}, 21(16):2894--2904, 2010.

\bibitem{berro2014}
Julien Berro and Thomas~D Pollard.
\newblock Local and global analysis of endocytic patch dynamics in fission
  yeast using a new "temporal superresolution" realignment method.
\newblock {\em Molecular biology of the cell}, 25(22):3501--3514, 2014.

\bibitem{kaksonen2005modular}
Marko Kaksonen, Christopher~P Toret, and David~G Drubin.
\newblock A modular design for the clathrin-and actin-mediated endocytosis
  machinery.
\newblock {\em Cell}, 123(2):305--320, 2005.

\bibitem{picco2015visualizing}
Andrea Picco, Markus Mund, Jonas Ries, Fran{\c{c}}ois N{\'e}d{\'e}lec, and
  Marko Kaksonen.
\newblock Visualizing the functional architecture of the endocytic machinery.
\newblock {\em eLife}, page e04535, 2015.

\bibitem{ewers:2009}
H.~Ewers, W.~R{\"o}mer, A.~Smith, K.~Bacia~S. Dmitrieff, W.~Chai, R.~Mancini,
  J.~Kartenbeck, V.~Chambon, L.~Berland, A.~Oppenheim, G.~Schwarzmann,
  T.~Feizi, P.~Schwille, P.~Sens, A.~Helenius, and L.~Johannes.
\newblock Gm1 structure determines sv40-induced membrane invagination and
  infection.
\newblock {\em Nat Cell Biol}, 12:11--18, 2009.

\bibitem{kukulski2012plasma}
Wanda Kukulski, Martin Schorb, Marko Kaksonen, and John~AG Briggs.
\newblock Plasma membrane reshaping during endocytosis is revealed by
  time-resolved electron tomography.
\newblock {\em Cell}, 150(3):508--520, 2012.

\bibitem{beauzamy2014flowers}
L{\'e}na Beauzamy, Naomi Nakayama, and Arezki Boudaoud.
\newblock Flowers under pressure: ins and outs of turgor regulation in
  development.
\newblock {\em Annals of botany}, page mcu187, 2014.

\bibitem{minc2009mechanical}
Nicolas Minc, Arezki Boudaoud, and Fred Chang.
\newblock Mechanical forces of fission yeast growth.
\newblock {\em Current Biology}, 19(13):1096--1101, 2009.

\bibitem{schaber2010biophysical}
J{\"o}rg Schaber, Miquel~{\`A}ngel Adrover, Emma Eriksson, Serge Pelet,
  Elzbieta Petelenz-Kurdziel, Dagmara Klein, Francesc Posas, Mattias
  Goks{\"o}r, Mathias Peter, Stefan Hohmann, et~al.
\newblock Biophysical properties of saccharomyces cerevisiae and their
  relationship with hog pathway activation.
\newblock {\em European Biophysics Journal}, 39(11):1547--1556, 2010.

\bibitem{de1996passive}
I~Martinez~de Mara{\~n}on, Pierre-Andr{\'e} Marechal, and Patrick Gervais.
\newblock Passive response of saccharomyces cerevisiae to osmotic shifts: cell
  volume variations depending on the physiological state.
\newblock {\em Biochemical and biophysical research communications},
  227:519--523, 1996.

\bibitem{carlsson2014force}
Anders~E Carlsson and Philip~V Bayly.
\newblock Force generation by endocytic actin patches in budding yeast.
\newblock {\em Biophysical journal}, 106(8):1596--1606, 2014.

\bibitem{kaksonen2006harnessing}
Marko Kaksonen, Christopher~P Toret, and David~G Drubin.
\newblock Harnessing actin dynamics for clathrin-mediated endocytosis.
\newblock {\em Nature reviews Molecular cell biology}, 7(6):404--414, 2006.

\bibitem{basu2014role}
Roshni Basu, Emilia~Laura Munteanu, and Fred Chang.
\newblock Role of turgor pressure in endocytosis in fission yeast.
\newblock {\em Molecular biology of the cell}, pages mbc--E13, 2014.

\bibitem{mcmahon:2005}
H.~T. McMahon and J.~L. Gallop.
\newblock Membrane curvature and mechanisms of dynamic cell membrane
  remodelling.
\newblock {\em Nature}, 438:590--596, 2005.

\bibitem{zimmerberg2006proteins}
Joshua Zimmerberg and Michael~M Kozlov.
\newblock How proteins produce cellular membrane curvature.
\newblock {\em Nature Reviews Molecular Cell Biology}, 7(1):9--19, 2006.

\bibitem{wigge1998amphiphysin}
Patrick Wigge and Harvey~T McMahon.
\newblock The amphiphysin family of proteins and their role in endocytosis at
  the synapse.
\newblock {\em Trends in neurosciences}, 21(8):339--344, 1998.

\bibitem{kirchhausen1986configuration}
T~Kirchhausen, SC~Harrison, and J~Heuser.
\newblock Configuration of clathrin trimers: evidence from electron microscopy.
\newblock {\em Journal of ultrastructure and molecular structure research},
  94(3):199--208, 1986.

\bibitem{peter2004bar}
Brian~J Peter, Helen~M Kent, Ian~G Mills, Yvonne Vallis, P~Jonathan~G Butler,
  Philip~R Evans, and Harvey~T McMahon.
\newblock Bar domains as sensors of membrane curvature: the amphiphysin bar
  structure.
\newblock {\em Science}, 303(5657):495--499, 2004.

\bibitem{reider2011endocytic}
Amanda Reider and Beverly Wendland.
\newblock Endocytic adaptors--social networking at the plasma membrane.
\newblock {\em Journal of cell science}, 124(10):1613--1622, 2011.

\bibitem{ford2002curvature}
Marijn~GJ Ford, Ian~G Mills, Brian~J Peter, Yvonne Vallis, Gerrit~JK Praefcke,
  Philip~R Evans, and Harvey~T McMahon.
\newblock Curvature of clathrin-coated pits driven by epsin.
\newblock {\em Nature}, 419(6905):361--366, 2002.

\bibitem{helfrich:1973}
W.~Helfrich.
\newblock Elastic properties of lipid bilayers---theory and possible
  experiments.
\newblock {\em Z. Naturforsch.}, C 28:693--703, 1973.

\bibitem{deuling1976curvature}
HJ~Deuling and W~Helfrich.
\newblock The curvature elasticity of fluid membranes: a catalogue of vesicle
  shapes.
\newblock {\em Journal de Physique}, 37(11):1335--1345, 1976.

\bibitem{wiese1992budding}
W~Wiese, W~Harbich, and W~Helfrich.
\newblock Budding of lipid bilayer vesicles and flat membranes.
\newblock {\em Journal of Physics: Condensed Matter}, 4(7):1647, 1992.

\bibitem{derenyi2002formation}
Imre Der{\'e}nyi, Frank J{\"u}licher, and Jacques Prost.
\newblock Formation and interaction of membrane tubes.
\newblock {\em Physical review letters}, 88(23):238101, 2002.

\bibitem{gao2005mechanics}
Huajian Gao, Wendong Shi, and Lambert~B Freund.
\newblock Mechanics of receptor-mediated endocytosis.
\newblock {\em Proceedings of the National Academy of Sciences of the United
  States of America}, 102(27):9469--9474, 2005.

\bibitem{nowak2008membrane}
Sarah~A Nowak and Tom Chou.
\newblock Membrane lipid segregation in endocytosis.
\newblock {\em Physical Review E}, 78(2):021908, 2008.

\bibitem{bovzivc2014direct}
Bojan Bo{\v{z}}i{\v{c}}, Jemal Guven, Pablo V{\'a}zquez-Montejo, and Sa{\v{s}}a
  Svetina.
\newblock Direct and remote constriction of membrane necks.
\newblock {\em Physical Review E}, 89(5):052701, 2014.

\bibitem{agrawal2010minimal}
Neeraj~J Agrawal, Jonathan Nukpezah, and Ravi Radhakrishnan.
\newblock Minimal mesoscale model for protein-mediated vesiculation in
  clathrin-dependent endocytosis.
\newblock {\em PLoS computational biology}, 6(9):e1000926, 2010.

\bibitem{Zhang2015508}
Tao Zhang, Rastko Sknepnek, M.J. Bowick, and J.M. Schwarz.
\newblock On the modeling of endocytosis in yeast.
\newblock {\em Biophysical Journal}, 108(3):508 -- 519, 2015.

\bibitem{walani2015endocytic}
Nikhil Walani, Jennifer Torres, and Ashutosh Agrawal.
\newblock Endocytic proteins drive vesicle growth via instability in high
  membrane tension environment.
\newblock {\em Proceedings of the National Academy of Sciences}, page
  201418491, 2015.

\bibitem{gustin1988mechanosensitive}
Michael~C Gustin, Xin-Liang Zhou, Boris Martinac, and Ching Kung.
\newblock A mechanosensitive ion channel in the yeast plasma membrane.
\newblock {\em Science}, 242(4879):762--765, 1988.

\bibitem{julicher1994shape}
Frank J{\"u}licher and Udo Seifert.
\newblock Shape equations for axisymmetric vesicles: a clarification.
\newblock {\em Physical Review E}, 49(5):4728, 1994.

\bibitem{stradalova2009furrow}
Vendula Str{\'a}dalov{\'a}, Wiebke Stahlschmidt, Guido Grossmann, Michaela
  Bla{\v{z}}{\'\i}kov{\'a}, Reinhard Rachel, Widmar Tanner, and Jan Malinsky.
\newblock Furrow-like invaginations of the yeast plasma membrane correspond to
  membrane compartment of can1.
\newblock {\em Journal of cell science}, 122(16):2887--2894, 2009.

\bibitem{kwok1981thermoelasticity}
R~Kwok and E~Evans.
\newblock Thermoelasticity of large lecithin bilayer vesicles.
\newblock {\em Biophysical journal}, 35(3):637--652, 1981.

\bibitem{morris2001cell}
CE~Morris and U~Homann.
\newblock Cell surface area regulation and membrane tension.
\newblock {\em The Journal of membrane biology}, 179(2):79--102, 2001.

\bibitem{evans:1990}
E.~Evans and W.~Rawicz.
\newblock Entropy-driven tension and bending elasticity in condensed-fluid
  membranes.
\newblock {\em Phys. Rev. Let.}, 64(17):2094--2097, Jan 1990.

\bibitem{cicuta2007diffusion}
Pietro Cicuta, Sarah~L Keller, and Sarah~L Veatch.
\newblock Diffusion of liquid domains in lipid bilayer membranes.
\newblock {\em The Journal of Physical Chemistry B}, 111(13):3328--3331, 2007.

\bibitem{jin2006measuring}
Albert~J Jin, Kondury Prasad, Paul~D Smith, Eileen~M Lafer, and Ralph Nossal.
\newblock Measuring the elasticity of clathrin-coated vesicles via atomic force
  microscopy.
\newblock {\em Biophysical journal}, 90(9):3333--3344, 2006.

\bibitem{cheng2007cryo}
Yifan Cheng, Werner Boll, Tomas Kirchhausen, Stephen~C Harrison, and Thomas
  Walz.
\newblock Cryo-electron tomography of clathrin-coated vesicles: structural
  implications for coat assembly.
\newblock {\em Journal of molecular biology}, 365(3):892--899, 2007.

\bibitem{bar1994instability}
Roy Bar-Ziv and Elisha Moses.
\newblock Instability and pearling states produced in tubular membranes by
  competition of curvature and tension.
\newblock {\em Physical review letters}, 73(10):1392, 1994.

\bibitem{kishimoto2011determinants}
Takuma Kishimoto, Yidi Sun, Christopher Buser, Jian Liu, Alph{\'e}e Michelot,
  and David~G Drubin.
\newblock Determinants of endocytic membrane geometry, stability, and scission.
\newblock {\em Proceedings of the National Academy of Sciences},
  108(44):E979--E988, 2011.

\bibitem{skruzny2012molecular}
Michal Skruzny, Thorsten Brach, Rodolfo Ciuffa, Sofia Rybina, Malte Wachsmuth,
  and Marko Kaksonen.
\newblock Molecular basis for coupling the plasma membrane to the actin
  cytoskeleton during clathrin-mediated endocytosis.
\newblock {\em Proceedings of the National Academy of Sciences},
  109(38):E2533--E2542, 2012.

\bibitem{godlee2013uncertain}
Camilla Godlee and Marko Kaksonen.
\newblock From uncertain beginnings: Initiation mechanisms of clathrin-mediated
  endocytosis.
\newblock {\em The Journal of cell biology}, 203(5):717--725, 2013.

\bibitem{mim2012structural}
Carsten Mim, Haosheng Cui, Joseph~A Gawronski-Salerno, Adam Frost, Edward
  Lyman, Gregory~A Voth, and Vinzenz~M Unger.
\newblock Structural basis of membrane bending by the n-bar protein endophilin.
\newblock {\em Cell}, 149(1):137--145, 2012.

\bibitem{lenz2009membrane}
Martin Lenz, Daniel~JG Crow, and Jean-Fran{\c{c}}ois Joanny.
\newblock Membrane buckling induced by curved filaments.
\newblock {\em Physical review letters}, 103(3):038101, 2009.

\bibitem{liu2006endocytic}
Jian Liu, Marko Kaksonen, David~G Drubin, and George Oster.
\newblock Endocytic vesicle scission by lipid phase boundary forces.
\newblock {\em Proceedings of the National Academy of Sciences},
  103(27):10277--10282, 2006.

\bibitem{liu2009mechanochemistry}
Jian Liu, Yidi Sun, David~G Drubin, and George~F Oster.
\newblock The mechanochemistry of endocytosis.
\newblock {\em PLoS biology}, 7(9), 2009.

\bibitem{tian2007line}
Aiwei Tian, Corinne Johnson, Wendy Wang, and Tobias Baumgart.
\newblock Line tension at fluid membrane domain boundaries measured by
  micropipette aspiration.
\newblock {\em Physical review letters}, 98(20):208102, 2007.

\bibitem{lipowsky2013spontaneous}
Reinhard Lipowsky.
\newblock Spontaneous tubulation of membranes and vesicles reveals membrane
  tension generated by spontaneous curvature.
\newblock {\em Faraday discussions}, 161:305--331, 2013.

\bibitem{saleem2015balance}
Mohammed Saleem, Sandrine Morlot, Annika Hohendahl, John Manzi, Martin Lenz,
  and Aur{\'e}lien Roux.
\newblock A balance between membrane elasticity and polymerization energy sets
  the shape of spherical clathrin coats.
\newblock {\em Nature communications}, 6, 2015.

\bibitem{boucrot2012membrane}
Emmanuel Boucrot, Adi Pick, Gamze Camdere, Nicole Liska, Emma Evergren,
  Harvey~T McMahon, and Michael~M Kozlov.
\newblock Membrane fission is promoted by insertion of amphipathic helices and
  is restricted by crescent bar domains.
\newblock {\em Cell}, 149(1):124--136, 2012.

\bibitem{grassart2014actin}
Alexandre Grassart, Aaron~T Cheng, Sun~Hae Hong, Fan Zhang, Nathan Zenzer,
  Yongmei Feng, David~M Briner, Gregory~D Davis, Dmitry Malkov, and David~G
  Drubin.
\newblock Actin and dynamin2 dynamics and interplay during clathrin-mediated
  endocytosis.
\newblock {\em The Journal of cell biology}, 205(5):721--735, 2014.

\bibitem{ferguson2012dynamin}
Shawn~M Ferguson and Pietro De~Camilli.
\newblock Dynamin, a membrane-remodelling gtpase.
\newblock {\em Nature reviews Molecular cell biology}, 13(2):75--88, 2012.

\bibitem{dommersnes1999n}
PG~Dommersnes and J-B Fournier.
\newblock N-body study of anisotropic membrane inclusions: Membrane mediated
  interactions and ordered aggregation.
\newblock {\em The European Physical Journal B-Condensed Matter and Complex
  Systems}, 12(1):9--12, 1999.

\bibitem{press2007numerical}
William~H Press.
\newblock {\em Numerical recipes 3rd edition: The art of scientific computing}.
\newblock Cambridge university press, 2007.

\bibitem{nelson1987fluctuations}
DR~Nelson and L~Peliti.
\newblock Fluctuations in membranes with crystalline and hexatic order.
\newblock {\em Journal de physique}, 48(7):1085--1092, 1987.

\bibitem{tsafrir2001pearling}
Ilan Tsafrir, Dror Sagi, Tamar Arzi, Marie-Alice Guedeau-Boudeville, Vidar
  Frette, Daniel Kandel, and Joel Stavans.
\newblock Pearling instabilities of membrane tubes with anchored polymers.
\newblock {\em Physical review letters}, 86(6):1138, 2001.

\bibitem{derenyi2007membrane}
I~Der{\'e}nyi, G~Koster, MM~Van~Duijn, A~Cz{\"o}vek, M~Dogterom, and J~Prost.
\newblock Membrane nanotubes.
\newblock pages 141--159, 2007.

\end{thebibliography}

\end{document}